\def\lesssim{\,\lower2truept\hbox{${<\atop\hbox{\raise4truept\hbox{$\
sim$}}}$}\,}
\def\gtrsim{\,\lower2truept\hbox{${>
\atop\hbox{\raise4truept\hbox{$\sim$}}}$}\,}

\def \SAIT #1 #2 {{\em Mem.\ Soc.\ Astron.\ It.\/} {\bf #1}, #2}
\def \MESS #1 #2 {{\em The Messenger\/} {#1}, #2}
\def \ASTRNACH #1 #2 {{ Astron. Nach.\/} { #1}, #2}
\def \AAP #1 #2 {{ A{\rm \&}A\/} {#1}, #2}
\def \AAL #1 #2 {{ A{\rm \&}A\/} {#1}, L#2}
\def \AAR #1 #2 {{ A{\rm \&}AR\/} {#1}, #2}
\def \AAS #1 #2 {{ A{\rm \&}AS\/} {#1}, #2}
\def \AJ #1 #2 {{ AJ\/} {#1}, #2}
\def \ANNREV #1 #2 {{ ARA{\rm \&}A\/} {#1},#2}
\def \APJ #1 #2 {{ ApJ\/} {#1}, #2}
\def \APJL #1 #2 {{ ApJ\/} {#1}, L#2}
\def \APJS #1 #2 {{ ApJS\/} {#1}, #2}
\def \APSS #1 #2 {{ Ap{\rm \&}SS\/} {#1}, #2}
\def \ASR #1 #2 {{ Adv. Space Res.\/} {#1}, #2}
\def \BAIC #1 #2 {{ Bull. Astron. Inst. Czechosl.\/} { #1}, #2}
\def \JSQRT #1 #2 {{ J. Quant. Spectrosc. Radiat. Transfer\/} {
#1}, #2}
\def \MN #1 #2 {{ MNRAS\/} { #1}, #2}
\def \MEM #1 #2 {{ Mem. R. Astr. Soc.\/} { #1}, #2}
\def \PLR #1 #2 {{ Phys. Lett. Rev.\/} { #1}, #2}
\def \PASJ #1 #2 {{ Publ. Astron. Soc. Japan\/} { #1}, #2}
\def \PASP #1 #2 {{ Publ. Astr. Soc. Pacific\/} { #1}, #2}
\def \NAT #1 #2 {{ Nat\/} { #1}, #2}
\def \ACTA #1 #2 {{ Acta Astron.\/} { #1}, #2}
\documentclass[]{aa}
\usepackage{graphicx}

    \def\smallskip{\vskip 6pt}
    
    \def\M12{${\rm M_{12}}$}

\begin{document}

\title{Far infrared and Radio emission in dusty starburst galaxies}

\author{A. Bressan\inst{1,2} \and L. Silva\inst{3} \and G. L. Granato\inst{1,2}}

\institute{INAF, Osservatorio Astronomico di Padova, Vicolo
Osservatorio 5, I-35122 Padova, Italy
            \and
   SISSA, Strada Costiera, I-34131 Trieste, Italy
               \and
   INAF, Osservatorio Astronomico di Trieste, Via Tiepolo 11, I-34131 Trieste, Italy}

\offprints{A. Bressan, bressan@pd.astro.it}

\date{Received / Accepted }

%%%%%%%%%%%%%%%%%%%%%%%%%%%%%%%%%%%%%%%%%%%%%%

\abstract{We revisit the nature of the far infrared (FIR)/Radio
correlation by means of the most recent models for star forming
galaxies, focusing in particular on the case of obscured
starbursts. We model the IR emission with our population
synthesis code, GRASIL (Silva et al. 1998). As for the radio
emission, we revisit the simple model of Condon \& Yin (1990). We
find that a tight FIR/Radio correlation is natural when the
synchrotron mechanism dominates over the inverse Compton, and the
electrons cooling time is shorter than the fading time of the
supernova (SN) rate. Observations indicate that both these
conditions are met in star forming galaxies, from normal spirals
to obscured starbursts. However, since the radio non thermal (NT)
emission is delayed, deviations are expected both in the early
phases of a starburst, when the radio thermal component
dominates, and in the post-starburst phase, when the bulk of the
NT component originates from less massive stars. We show that
this delay allows the analysis of obscured starbursts with a time
resolution of a few tens of Myrs, unreachable with other star
formation (SF) indicators. We suggest to complement the analysis
of the deviations from the FIR/Radio correlation with the radio
slope (q--Radio slope diagram) to obtain characteristic
parameters of the burst, e.g.\ its intensity, age and fading time
scale. The analysis of a sample of compact ULIRGs shows that they
are intense but transient starbursts, to which one should not
apply usual SF indicators devised for constant SF rates. We also
discuss the possibility of using the q--radio slope diagram to
asses the presence of obscured AGN. A firm prediction of the
models is an apparent radio excess during the post-starburst
phase, which seems to be typical of a class of star forming
galaxies in rich cluster cores. Finally we discuss how deviations
from the correlation, due to the evolutionary status of the
starburst, affect the technique of photometric redshift
determination widely used for the high-z sources.

\keywords{
-- Interstellar medium: dust extinction
-- Galaxies: stellar content
-- Infrared: galaxies
-- Radio continuum: galaxies}
}

\titlerunning{FIR and Radio emission in dusty starbursts}
\authorrunning{A. Bressan et al.}

\maketitle

\section{Introduction}

In recent years the study of starburst galaxies has become a very
popular subject because of its intimate connection with the
global star formation history of the Universe. On one side high
redshift observations in the optical bands probe rest frame
spectral regions that are highly affected by even tiny amounts of
ongoing star formation and dust extinction. On the other,
theoretical models following the paradigm of the hierarchical
clustering scenario predict that merging induced star formation
should be highly enhanced in the past. Current estimates of the
star formation rate (SFR) of the Universe have thus been
interpreted on the basis of our understanding of local analogous
galaxies, in particular through UV continuum and optical line
emission. However in local  starbursts a significant fraction of
the ongoing star formation may be hidden to UV and optical
estimators. In fact, though starburst galaxies were initially
selected for the prominence of their optical emission lines, it
appears that this criterion excludes other actively star forming
objects and possibly limits our understanding to a small phase of
their evolution. After the IRAS satellite, it became clear that
SF is also highly enhanced in very and ultra luminous infrared
galaxies that are otherwise highly attenuated in the optical. 
With space densities similar to those of quasars (Soifer et al. 1986) and total
infrared luminosities spanning the range  10$^{11} $--10$^{12}$L$_{\odot}$ and
above 10$^{12}$L$_{\odot}$, respectively,  
Luminous  and Ultraluminous Infrared
galaxies (LIRGs and ULIRGs) are
the most luminous objects in the local Universe.
Evidence of the important role played by dust
reprocessing was provided by the detection of a diffuse FIR
background whose high intensity (equal to or higher than that of
the optical, e.g. Hauser et al. 1998) implies that these galaxies
are undergoing intense star formation activity (Puget et al.
1996; Dwek  et al. 1998). Furthermore, the advent of the Infrared
Space Observatory (ISO), in combination with the availability of
new ground facilities such as SCUBA on the JCMT, have discovered
numerous high-z galaxies with enhanced IR emission (e.g. Elbaz et
al. 1999; Smail et al. 2000; Barger et al. 2000).

Silva et al.\ (1998) have first introduced the concept of
age-selective obscuration, to explain the features of the
observed spectral energy distribution (SED) of star forming
galaxies, from normal spirals to dust obscured starbursts, from
the UV to the sub-mm. In this model young stars are supposed to
originate within molecular clouds and correspondingly their light
is attenuated more than that of older stars, that already got rid
of their parental cloud. The UV light in many starbursts is thus
dominated by the older stars rather than by the younger
populations. With the same assumptions, Granato et al.\ (2000)
reproduced the observable properties of local galaxies (in
particular the IRAS luminosity function), working within the
context of structure formation through hierarchical clustering,
which has successfully confronted a wide range of observations on
large scale structure and microwave background anisotropies. They
showed that the concept of age-selective obscuration could
explain the difference between the galactic {\it extinction} law
and the {\it attenuation} law observed in starburst galaxies
(Calzetti, Kinney, \& Storchi-Bergmann 1994).

Poggianti, Bressan, \& Franceschini (2001) have recently
investigated the optical spectra of very luminous infrared
galaxies to constrain the recent history of SF
and the dust extinction characteristic of various
stellar populations. They have found that the most plausible
explanation for their unusual combination of strong H$\delta$
absorption and moderate [OII] emission is again age-selective
extinction. Indeed HII regions (wherein the [OII] emission
originates) are highly embedded and thus are affected by a
greater extinction compared to the older stellar populations
which are responsible for the Balmer absorption. Under standard
assumptions for the IMF, the SFR derived from the
fit of the optical spectrum (continuum, absorption and emission
lines) may account for a small fraction of the FIR emission.
Moreover, even complemented with the information on the FIR flux,
the optical-UV spectrum is not enough to identify univocal
solutions. Further evidence along this direction is provided by
recent observations of UV properties of ULIRGs (Goldader et al.\
2002).

These studies underline an intrinsic difficulty of evaluating the
properties of massive starbursts only from their UV, optical and
even NIR properties and the natural way out from this impasse
seems provided by studies at longer wavelengths. The capability
of FIR and radio spectral regions to reveal otherwise hidden
complex phenomena in star forming galaxies, is testified by the
existence of a "miraculous" correlation between their properties
in these spectral windows. The FIR/Radio correlation is locally
well established over a significant range of luminosity, from
normal spirals to the most extreme ULIRGs and its small scatter
states the universal proportions with which energy is radiated
away at IR and radio wavelengths. In spite of its obscure nature,
its utility appears in several aspects, beside being a firm tool
within the manifold of star formation indicators. For instance the
validity of the FIR/Radio correlation has been recently confirmed
up to redshift $\simeq 1.3$ (Garret 2001) and it is widely
extrapolated much beyond, to estimate the redshift of more
distant objects (e.g. Carilli \& Yun 2000). Also, deviations from
the correlation observed toward the central regions of rich
cluster of galaxies, where a significant fraction of star forming
galaxies show a radio excess, are used to trace the effect of the
hot intracluster medium on their galactic magnetic field (Gavazzi
\& Jaffe 1986; Miller \& Owen 2001).

So far there have been many attempts to explain the FIR/Radio
correlation but all have soon or late invoked a fine tuning of the
relevant physical properties, such as the intensity of the
radiative and magnetic energy density (e.g.\ Lisenfeld, Volk \&
Xu 1996). In this paper we revisit this correlation by combining
our spectrophotometric code GRASIL, particularly suited to the
study of the IR properties of dusty galaxies, with a new model of
radio emission. The latter essentially follows the recipes by
Condon (1992), but after a careful assessment of the validity of
one of its basic assumptions, namely the proportionality between
the non thermal (NT) radio emission and the core-collapse
supernova (CCSN) rate. This fact renders the FIR/Radio correlation
so robust and we provide, for the first time, a simple explanation
of its universality. For the same reason we show that deviations
are to be expected during and soon after the starburst episode,
and we suggest that they can help in constraining the star
formation history of these galaxies, something that cannot be
done with optical, NIR and even FIR observations.

In Sect.~\ref{sfir} we briefly describe our population synthesis
code for dusty galaxies. Sect.~\ref{sradio} is devoted to the new
model of radio emission. In Sect.~\ref{scalib} we describe our
calibration of the NT radio emission model and obtain new
relations for the SFR against radio emission  for the case of
quiescent galaxies. In Sect.~\ref{sq} we analyse infrared and
radio properties of starburst galaxies. We show that the
different fading times of FIR and radio emissions may be used to
reach a time resolution of a few tens of Myr, which is impossible
resting only on the UV-FIR. 
In Sect.~\ref{squ} we introduce a new diagnostic
tool, the FIR/Radio (q) vs radio spectral slope diagram, which
potentially allows the determination of the evolutionary status
of a starburst in absence of a good radio spectral coverage. 
We examine the location of an observed sample of compact ULIRGs
and discuss whether this diagram may also
provide a quantitative estimate of the threshold between AGN and
star formation powered ULIRGs.
In Sect.~\ref{spsbt} we analyse
the evolution in the post-starburst phase and suggest that radio
excess is actually an indication of the occurrence of this phase
rather than an environmental effect. In Sect.~\ref{shz} we
discuss the impact of these new findings on the determination of
the redshift of SCUBA sources, a method that relies on the
FIR/Radio correlation. Sect.~\ref{sconc} is devoted to our
conclusions.

\section{Infrared Emission}
\label{sfir}

Infrared emission is calculated with GRASIL, a code
designed to perform population synthesis in presence of
dust (Silva et el. 1998).
In brief, the star formation history, the metal enrichment and
the current gas fraction are  provided by a chemical evolution
code. Stars and dust are distributed either in a disk or in a
bulge or both, including high density clumps, the molecular
clouds. Young stars are supposed to originate within molecular
clouds and to leave them in a characteristic time scale
t$_{esc}$. The volume emissivity is computed by considering the
light of young stars, absorbed by the molecular clouds, and the
light of older populations, both propagated through the diffuse
dust component. The code has been thoroughly tested against
observations (e.g.\ Silva et al.\ 1998; Granato et al.\ 2000).

For a given star formation history, gas fraction and
metallicity, one of the parameters that largely affects the FIR
emission is the escape time t$_{esc}$. Silva et al.\ (1998) and
Granato et al.\ (2000) have shown that the UV, optical and FIR
properties of local spirals are well reproduced  with a typical
t$_{esc} \simeq 3$ Myr. Normal star forming regions are also
characterized by a moderate visual attenuation A$_V\simeq 1$ mag.
In the case of dusty starbursts, the slope of the UV continuum
(Meurer et al. 1999) indicates large obscuration times. Silva et
al.\ (1998) and Granato et al.\ (2000) were able to reproduce the
SEDs, from the UV to the FIR, with a t$_{esc}$ in excess of several Myr and
a compact geometry with
a characteristic radius of a fraction of a Kpc. Similar
obscuration times seem to be required to interpret
the optical spectra of very luminous infrared galaxies (Poggianti
et al. 2001).

We have considered as representative of a normal spiral a model
with the parameters shown in Table~\ref{chem}. We adopt a Schmidt
type star formation law, i.e. SFR=$\nu \, \mbox{M}_g^{k}$, with
infall of gas with primordial composition in a time scale
t$_{inf}$, and a Salpeter initial mass function (IMF) (slope
x=1.35 in mass), from 0.15 M$_\odot$ to 120 M$_\odot$. For
starburst galaxies we superimpose an exponentially decreasing
burst ($t_b$=10, 15, 25, 50 Myr) of star formation at an epoch of
11.95 Gyr, with the same IMF. The mass of stars formed during the
burst is set to 10\% of the disk mass. Our main conclusions are
not affected by details in these choices.

\begin{table} \caption{Parameters for a disk galaxy model. See
text for details.}
\begin{tabular}{cccccc}
\hline
$\nu$  & k & t$_{inf}$ & x &
M$_{Low}$ & M$_{UP}$ \\
\hline 0.5 Gyr$^{-1}$ & 1 & 9~Gyr &1.35 & 0.15 M$_\odot$ &120
M$_\odot$ \\
\hline
\end{tabular}
\label{chem}
\end{table}

\section{Radio emission}
\label{sradio}

A clear picture of radio emission from normal galaxies is still
missing, particularly for the often dominant non thermal
component. Indeed, it is well known that the intensity of the
thermal component is tightly related to the number of H ionizing
photons, Q(H), and scales as $\approx \nu^{-0.1}$
 (Rubin, 1968).
Computing HII
region models with CLOUDY (Ferland 1996), for different 
mass (10$^4$M$_\odot$ to 10$^5$M$_\odot$), 
metallicity (Z = 0.008, 0.02 and 0.05) and 
age (1 to 10 Myr) of the ionizing cluster, and 
for electron densities of 10$^1$ to 10$^3$ cm$^{-3}$
and inner radii of the nebula of 10 and 100 pc,
we have obtained the
following average relation at 1.49 GHz (see Panuzzo et al. 2002 for
details of the inclusion of nebular emission in GRASIL):
\begin{equation} \frac{L^T_{\nu }}
{\mbox{erg/s/Hz}}\simeq \frac{Q(H)}{5.495\times 10^{25}}
\left(\frac{T_{e} }{10^{4} \mbox{K}}\right)^{0.45}\left(\frac{\nu
}{1.49\mbox{GHz}}\right)^{-0.1} \label{tt}
\end{equation}
It is then straightforward to obtain Q(H) and the intensity of
thermal radio emission from simple stellar populations.

On the other hand, little is known about the source of the non
thermal emission which, in normal star forming galaxies may
account 90\% of the radio emission (Condon 1992). Observations
indicate that FIR and radio emission are strongly correlated over
a wide range of IR luminosities. At 1.49 GHz (Sanders \& Mirabel
1996):
\begin{equation}
q=\log \frac{\mbox{F}_{\mbox{FIR}}/(3.75\times 10^{12}\mbox{Hz})}{F_\nu
(1.49\mbox{GHz})/(\mbox{W m}^{-2}\mbox{Hz}^{-1})}\simeq 2.35\pm 0.2
\label{qeq}
\end{equation}
where F$_{FIR}=1.26 \, 10^{-14}(2.58\, S_{60\mu{\rm m}} +
S_{100\mu{\rm m}})$ $\mbox{W m}^{-2}$, with $S_{60}$ and $S_{100}$
in Jy. This correlation suggests that NT emission is
related to the recent star formation and the most likely
mechanism is synchrotron emission from relativistic electrons
accelerated into the shocked interstellar medium, following
CCSN explosions. But the poor knowledge of the
accelerating mechanism hinders any quantitative prediction of
this  phenomenon which can only  be formulated in an empirical
way (Condon 1992). The observed average luminosity per supernova
event can be estimated with the ratio between NT radio emission
and CCSN rate ($\nu _{CCSN}$) in our Galaxy (e.g.\ Condon \& Yin
1990). With $\nu_{CCSN}\simeq~0.015$ (Turatto,
private communication) and after converting
$L_{0.4\mbox{GHz}}~\simeq 6.1\times~10^{21} \mbox{W Hz}^{-1}$
(Berkhuijsen, 1984) to 1.49 GHz by assuming a radio slope of
$\alpha \equiv -\frac{d\,\log S_\nu}{d\, \log \nu}$=0.8, we obtain
\begin{equation}
E^{NT}_{1.49}=\frac{L^{NT}_{1.49\mbox{GHz}}/(10^{30}\mbox{erg~s}^{-1}~\mbox{Hz}^{-1})}{\nu_{CCSN}/\mbox{yr}^{-1}}
\simeq 1.44
\label{ntgal}
\end{equation}

It is still a matter of debate  why the FIR/radio correlation is
unaffected by the dependence of the electrons lifetime and
luminosity on the magnetic and radiation density fields, which
may change significantly  in different environments.

Lisenfeld et al.\ (1996) claimed that, in normal star forming
galaxies,  the existence of the FIR/Radio correlation requires a
correlation between the radiation and magnetic energy density
fields. However the radiation field changes dramatically in
starburst galaxies, and there must be a significant fine tuning
between the parameters regulating the intensity of the two fields,
because they must scale in such a way that they give rise to the
FIR/Radio correlation and, at the same time, they  must prevent
inverse Compton to dominate over synchrotron losses.

As a possible way out of this conundrum, we  suggest here  that a
FIR/Radio correlation originates because synchrotron electron lifetimes
are shorter than the fading time of the CCSN rate.
%This may be the case of normal spirals, where the star formation rate
%remains constant over a significant time, as well as among
%powerfull starbursts, where the presence of synchrotron
%emission in spite of an enhanced radiation field, suggests
%high magnetic fileds and, consequntly, short electron lifetimes.
Assuming that cosmic ray electrons are injected during the
adiabatic phase of SN explosions (t$\leq$10$^4$--10$^5$yr), i.e.\
in a characteristic time scale which is short compared to the
star formation time scale, the bolometric synchrotron luminosity
at an epoch T is given by
\begin{equation}
L^{NT}=\int_0^{min(T,\tau_s^{el})}\nu_{CCSN}(T-t) \, l^{NT}(t)dt
\label{ntsfr}
\end{equation}
where t is the lookback time, 
$l^{NT}(t)$ is the NT luminosity of the injected electrons
after a time t has elapsed, T is the age of the galaxy and
$\tau_s^{el}$ is the lifetime of electrons against synchrotron
losses and depends on the intensity of the magnetic field. For
normal spiral galaxies $\tau_s^{el}$ ($<<$ T) may be as large  as several
10$^7$ yr but, since the star formation rate is almost constant
over the last Gyr, the above
integral becomes
\begin{equation}
L^{NT}=\nu_{CCSN}\int_0^{\tau_s^{el}}\frac{dE}{dt}dt=\nu_{CCSN}E^{el}
\label{eel}
\end{equation}
where we put $l^{NT}(t)=dE/dt$. Thus the bolometric NT luminosity
scales linearly with the SN rate and the proportionality constant
is the injected energy of the electrons per SN (E$^{el}$).

In the case of starburst galaxies the SN rate cannot be
considered constant over such a large time scale. However also
$\tau_s^{el}$ must be  much shorter: to avoid significant losses
from inverse Compton on the intense stellar radiation field
$\tau_s^{el} <<$1-0.1 Myr (Condon 1992). Thus we may still make
use of the approximation in Eq.~\ref{eel}, because we may
consider $\nu_{CCSN}$ almost constant over such a small time
scale.

In brief, the NT radio luminosity of a galaxy is proportional to
the integral of the synchrotron power over the electron lifetime, and an
increase of the former in a larger magnetic field is compensated by a
shortening of the latter.

Since both the SN rate and the FIR emission are strictly related
to the recent star formation rate, our justification of the
validity of Eq.~\ref{eel} in very different environments,
explains why the FIR/Radio correlation is so robust. No fine
tuning is necessary, apart from the requirement that the magnetic
field is large in starburst galaxies, which, by itself, is an
independent observational fact.

The considerations above apply to the bolometric radio luminosity,
but what is actually measured is the specific luminosity
$L^{NT}_{\nu}$. However, under plausible assumptions, also 
$L^{NT}_{\nu }$
scales linearly with the SN rate, with the only dependence on
environmental conditions being an almost vanishing one on the
magnetic field. This is a consequence of the observed spectral
index of the NT radio emission.

Indeed, the diffusion-loss equation (e.g. Longair 1994) for the
time evolution of the number density of relativistic electrons
per unit energy interval $N(E,t)$ is
\begin{equation}
\frac{dN(E,t)}{dt} =\frac{d}{dE} \lbrack
b(E)N(E,t)]+Q(E,t)+D\nabla ^{2} N(E,t) \label{eqdif}
\end{equation}
where $b(E)$ is the rate at which particles lose energy, $Q(E,t)$
is the rate at which electrons are injected in the system per
unit volume, time and energy interval (the source term) and $D$
is the diffusion coefficient. We assume that the distribution of
sources of fresh electrons (i.e. the SNae) is sufficiently
uniform and extended to make negligible the diffusion term, and
that each SN event injects high energy electrons with a power law
energy spectrum $\propto E^{-p} $, so that the source term can be
written as
\[
Q(E,t)=k\frac{\nu _{CCSN} (t)}{V} E^{-p}
\]
where $k$ and $p$ are constants (depending on the detailed physics
of SN explosion) and $V$ is the volume of the system. Also, we
are interested in stationary solutions $dN/dt=0$, that is we
consider electron energies at which the lifetime against
radiative losses is short compared to the typical time scale for
variations of $\nu_{CCSN}(t)$. The solution of Eq. \ref{eqdif}
is then simply
\[
N(E)=\frac{k\nu_{CCSN} E^{-(p-1)} }{V(p-1)b(E)}
\]
where $b(E)$ is the rate at which particles lose energy. If the
dominant loss mechanism is synchrotron radiation,
\[
b(E)=-\frac{dE}{dt} _{synch} =\frac{4}{3} \sigma _{T} c\gamma
^{2} \frac{B^{2} }{8\pi }
\]
In this case
\[
N(E)=\frac{Ak\nu_{CCSN} }{V} \frac{E^{-(p+1)} }{B^{2} }
\]
where the multiplicative factor $A$ depends only on $p$ and
fundamental constants: electrons with a sufficiently short
lifetime have a power law energy distribution, with an index
steeper by 1 than the injected distribution.

On the other hand, the synchrotron luminosity of an optically
thin source with a random magnetic field and electrons having a
power law energy distribution $N(E)=CE^{-q} $ is
\[
L^{NT}_{\nu } =j_{\nu ,synch} V=C\,D(q)\/\,V\,B^{\frac{q+1}{2} } \nu
^{-\frac{q-1}{2} }
\]
$D(q)$ depends only on $q$ and fundamental constants. This
equation with $C=Ak\nu_{CCSN} /VB^{2} $ and $\alpha =(q-1)/2$
becomes
\[
L^{NT}_{\nu } =\nu_{CCSN} \left( A\,kD(p)\/\,\,B^{\alpha -1} \nu
^{-\alpha } \right)
\]
which shows that also the specific luminosity is proportional to
the SN rate, with a proportionality factor which, given the
observed spectral index of radio emission $\alpha \approx 1$, has
only a very weak dependence on the magnetic field. It may be
useful to note explicitly that the above formula integrated over
$\nu$ yields again the result expressed by Eq. \ref{eel}, namely
that the bolometric NT luminosity does not depend on $B$. This is
because the limit of integration in frequency are those
corresponding to the limits in energy within which the power law
electron distribution applies, and the link between the two is
$\nu \propto B\times E$.

The crucial assumptions in this derivation are that the SN rate
is constant over time scales of the order of the lifetimes of
relevant electrons, and that the dominant electron energy loss is
synchrotron emission, as already discussed.

\subsection{The contribution of Radio Supernova Remnants}

Before discussing how to calibrate Eq.~\ref{eel}, we consider the
possible sources of non thermal radio emission in our own Galaxy.

We have checked that, among the identified galactic sources, only
radio Supernova Remnants (SNR) may provide a significant
contribution to the NT radio emission. Other sources like
the pulsars themselves and the bubbles of NT radio
emission associated with X-ray binaries provide a negligible
contribution.
However SNRs, while being the obvious most appealing sources,
cannot be responsible of the  bulk of NT radio emission in normal
galaxies for two reasons. First, their spectrum has a characteristic
radio slope $L_{\nu } \propto \nu ^{-\alpha }$ with a broad range of
$\alpha$ between  0.2 and 0.5  (Gordon et al.\ 1999) which is, on the
average, less than that characteristic of normal galaxies
($\alpha\simeq$0.8).
Second, one may easily show that they cannot supply more than
about 5\% of the total NT luminosity, or, equivalently, the SN
rate needed to reproduce the NT radio emission of the
Galaxy is more than one order of magnitude larger than observed
(see e.g.\ Condon 1992 and references therein).

In fact, because the typical lifetime of a SNR is of a few
10$^4$yr, we may estimate the contribution of a {\sl population}
of SNRs originated by an instantaneous burst of star formation, by
applying the fuel consumption theorem of post main sequence
evolutionary phases (Renzini \& Buzzoni 1986). Indeed short
evolutionary phases beyond the main sequence provide an
integrated luminosity
\begin{equation}
L^{SSP}=\phi(m_d) \frac{dm_{d}}{dt} \int_{t_{beg}}^{t_{end}} l_d(t) dt
\label{ls}
\end{equation}
where $m_d$ is the dying mass ($\simeq~M_{\rm Turn-off}$),
$l_d(t)$ is the luminosity evolution within the phase,
$\phi(m_d)$ is the IMF, $\frac{dm_{d}}{dt}$ is the time derivative
of the dying mass, and the integral refers to the particular
phase. Altogether the quantity $\phi(m_d)\frac{dm_{d}}{dt}$
provides the evolutionary flux, namely the rate of dying stars of
the given population.

The luminosity evolution of a single SNR may be derived by
combining the observed surface brightness--diameter ($\Sigma -D$)
relation at 408 MHz
\begin{equation}
\Sigma ~(\mbox{W m}^{-2} \mbox{Hz}^{-1} \mbox{sr}^{-1})\simeq
10^{-15} \, D_{\rm pc}^{-3}
\end{equation}
with the time evolution of the linear diameter of a SNR (Clark \& Caswell 1976)
\begin{equation}
D_{\rm pc}\simeq 0.43 \, E_{50}^{1/5}n^{-1/5}t^{2/5}
\end{equation}
It follows that
\begin{equation}
l(\mbox{erg s}^{-1} \mbox{Hz}^{-1}) \simeq 3.55 \times 10^{26} \,
D_{\rm pc}^{-1}
\end{equation}
and the evolution of the luminosity of a SNR (at 1.49 GHz, assuming
a slope $\alpha$=0.3) is
(see e.g.\ Condon 1992)
\begin{equation}
\frac{l_{1.49}(t)}{\mbox{erg/s/Hz}}\simeq
10^{26}\,E_{50}^{-1/5} n^{1/5} t^{-2/5}
\label{lradt}
\end{equation}

The integrated contribution of the population of SNRs from
an instantaneous burst of star formation, at an age t, is then
\begin{equation}
\frac{L^{SSP}_{1.49}}{\mbox{erg/s/Hz}}\simeq 1.7\times 10^{26}\,E_{50}^{-1/5}n^{1/5}
\tau_{c}^{3/5}\phi (m_{d})\frac{dm_{d}}{dt}
\end{equation}
where $E_{50}$ is the SN blast energy in units of $10^{50}$ erg,
$n$ (cm$^{-3}$) is the ambient particle 
density and $\tau_{c}$ is the lifetime
of the SNR in yr. Note that $m_d$ is a function of time, but it
is almost constant during the considered phase.

The lifetime of a SNR is usually associated with its
adiabatic phase (Condon 1992)
\begin{equation}
\tau_{c}\simeq 2\times 10^{4}\,E_{50}^{4/17}n^{-9/17} \ \mbox{yr}
\end{equation}
thus
%\begin{equation}
%L^{SSP}_{1.49}(t)=0.\,\allowbreak 75\times 10^{29}\,n^{-2/17}E_{50}^{-1/17}\phi(m_{d})
%\frac{dm_{d}}{dt}.
%\end{equation}
\begin{equation}
\frac{L^{SSP}_{1.49}}{\mbox{erg/s/Hz}} \simeq 0.\,\allowbreak 6\times 10^{29}\,n^{-2/17}E_{50}^{-1/17}\phi(m_{d})
\frac{dm_{d}}{dt}.
\end{equation}
Integrating over the past SFR, $\varphi(t)$, until we still get
CCSN events (t~$\leq$~t$_{CCSN}$) and after defining the current
supernova rate
\begin{equation}
\nu_{CCSN}(\mbox{yr}^{-1})=\int_0^{t_{CCSN}}\phi (m_{d})
\frac{dm_{d}}{dt}\varphi(t)dt.
\end{equation}
we obtain for the average SNR non thermal luminosity per supernova event
\begin{equation}
E_{1.49}^{SNR}= \frac{L^{SNR}_{1.49}/(10^{30}{\rm
erg/s/Hz})}{\nu_{CCSN}/{\rm yr}^{-1}} \simeq 0.06 \,
n^{-2/17}E_{50}^{-1/17} \label{qsnr}
\end{equation}
This result, independent from the assumed IMF and star
formation history, is also not very sensitive to the 
environment. It clearly shows that the contribution of SNRs is
much less than required by Eq.~\ref{ntgal} for plausible values
of $n$ and $E_{50}$ (Condon 1992). In principle the lifetime of a
SNR is longer than the adiabatic time and could also depend on
the ambient density. Kafatos et al.\ (1980) quote for a hot
cavity SNR $\tau_{SNR}=1.4\times ~10^{5}$ yr, at which time the
linear diameter is about $100$ pc and the velocity of the shock
front is $300$ km/s, still capable of maintaining a shock
temperature of about $10^{6}$ K. However this lifetime is still
one order of magnitude less than that required by the observed NT
luminosity ($L\propto \tau ^{3/5}$) and, more important, when the
diameter is larger than about $20$ pc the observed $\Sigma -D$
relation steepens considerably ($\Sigma \propto D^{-10}$) so
that, in any case, the contribution beyond $\tau_{c}$ is
negligible.

\section{Calibration of the NT radio emission}
\label{scalib}

In summary in modelling the properties of the NT radio luminosity
in our Galaxy we have found that the only non negligible discrete
sources SNRs can provide at most 6\% of the NT radio luminosity
(e.g.\ Condon 1992). Following Condon \& Yin (1990) we have thus
calibrated Eq.~\ref{eel} (after accounting for the small
contribution of SNRs) against the SN rate and synchrotron
luminosity of our Galaxy. We assume for the NT radio emission
\begin{eqnarray}
&&\frac{L^{NT}(\nu)}{10^{30}\mbox{erg s}^{-1} \mbox{Hz}^{-1}}=
\\ \nonumber &&\left(E_{1.49}^{SNR}(\frac{\nu}{1.49})^{-0.5} +
E_{1.49}^{el}(\frac{\nu}{1.49})^{-\alpha} \right) \times
\frac{\nu_{CCSN}}{\mbox{yr}^{-1}} \label{ntt}
\end{eqnarray}
Since the NT contribution of SNRs has been evaluated in
Eq.~\ref{qsnr} ($E_{1.49}^{SNR}\simeq$ 0.06) then Eq.~\ref{ntgal}
requires that $E_{1.49}^{el}\simeq 1.38$. In order to reproduce
the observed average slope of the NT radio emission in normal
spiral galaxies ($\simeq$0.8), we have adopted $\alpha\simeq$0.9.

An independent check of this calibration is provided by the
observed ratio between the FIR and radio emission in normal
spirals (Eq.~\ref{qeq}). We have evolved several models with
GRASIL for different values of the critical parameters for the
chemical evolution (different age, gas infall time scale and star
formation efficiency), the escape time from molecular clouds (2
Myr and 3 Myr, Granato et al. 2000) and the scale length of the
dust distribution (3 Kpc and 4 Kpc ). The values of q turn out to
be quite independent from the adopted parameters and cluster
around q=2.35, in excellent agreement with the observations. We
emphasize that, in our models, the SFR depends linearly on the
gas fraction (Schmidt law), while the FIR emission depends on the
SFR, gas fraction and metallicity, and the SN rate on the recent
SFR. The consistency between FIR emission, radio emission and
supernova rate, is thus remarkable and should be considered as a
successful test of the model. We  have also obtained the
following calibrations between radio emission and star formation
rate:
\[
SFR({\rm M}_\odot/{\rm yr}) = 2.8 \, 10^{-28} \, L(8.4{\rm
GHz})/({\rm erg/s/Hz})
\]
\[
SFR({\rm M}_\odot/{\rm yr}) = 7.5 \, 10^{-29} \, L(1.49{\rm
GHz})/({\rm erg/s/Hz})
\]
These relations hold only for the case of a constant SFR for more
than 100 Myr and are in good agreement with those obtained by
Carilli (2000). It is worth noticing that with a SN rate of 0.023
-- the value adopted by Condon 1992 -- we would have obtained a
SFR calibration higher by about 50\%.

\begin{table}
\caption{Starburst parameters for the SEDs in Figg.~\ref{m82} and \ref{arp220}}
\begin{tabular}{lcccc}
\hline
Case & Age   & $t_b$ & Current SFR & Burst Mass \\
     & (Myr) &  (Myr) &(M$_\odot$/yr)&(M/M$_{tot}$) \\
\hline
M82  & &  &  &\\
a {\it dotted} &45 & 50 & 3.41 & 0.011\\
b {\it solid} &24 & 8  & 2.37 & 0.011 \\
\hline
ARP220  & &  &  &\\
a {\it solid} & 50 & 50 & 271 & 0.121\\
b {\it dotted} & 40 & 30 & 280 & 0.121 \\
c {\it dashed} & 25 & 8  & 149 & 0.121 \\
\hline
\end{tabular}
\label{tabir}
\end{table}

Before concluding this section it is worth commenting on our
choice of the lower limit of the initial mass giving rise to type
II SNe. It is commonly assumed that it corresponds to an initial
mass of about 8 M$_\odot$ for standard convection while it lowers
to about 6 M$_\odot$ for models with convective overshoot.
However, recent investigations on the evolution between 8 and 10
M$_\odot$ (without overshoot) indicate that the final fate of
stars born in this mass range is that of a white dwarf, instead
of an electron capture SN (Ritossa, Garcia-Berro \& Iben 1996).
The same fate is expected in the mass range 6 to 8 M$_\odot$ if
the overshoot scheme is adopted (Portinari, Chiosi, \& Bressan
1998). We have thus assumed that CCSN are produced in stars with
mass M$\geq$8 M$\odot$ and for ages younger than
t$_{CCSN}\simeq$50Myr.

\section{FIR/Radio properties of starburst models}
\label{sq}

\begin{figure}
\centering
\includegraphics[width=8truecm]{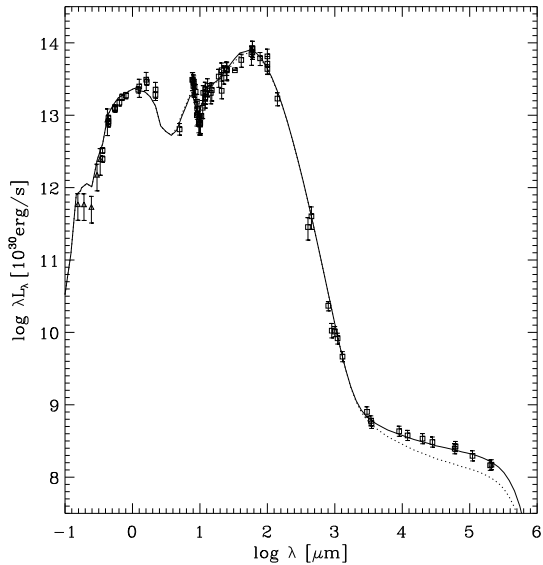}
\caption{M82: two models with different age and e-folding time scale (Tab. \ref{tabir}),
fit the observed SED from UV to the sub-mm. Inclusion of radio wavelengths is necessary
to disentangle the two cases. For details on the models see Silva et al.\
(1998).}
\label{m82}
\end{figure}
\begin{figure}
\centering
\includegraphics[width=8truecm]{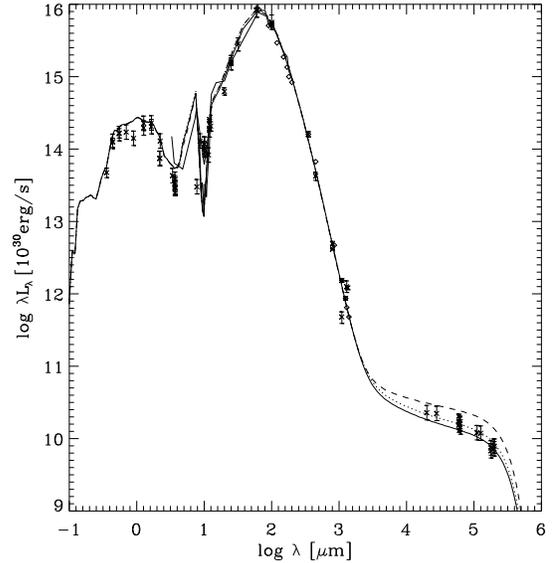}
\caption{Same as in Fig.~\ref{m82}, for three models for ARP220.}
\label{arp220}
\end{figure}
In this section we examine the FIR and radio properties of our new models and
compare them with two well studied local starbursts, M82 and ARP220. To obtain a
realistic starburst model, we added an exponentially decreasing burst of star
formation to the underlying secular disk evolution of the spirals models used in
the previous section. Table~\ref{tabir} summarizes the burst parameters in few
adopted models. Notice that in all models of the same galaxy, the total mass of gas
converted into stars is the same (column $M/M_{tot}$).
The corresponding GRASIL parameters have been selected according to
Silva et al. (1998) and, in particular, the adopted obscuration time must be
significantly larger than that characteristic of normal galaxies.
 
Fig.~\ref{m82} shows two fits to the SED of M82,
differing only in the age (45 and 24 Myr)
and e--folding time (50 and 8 Myr) of the burst. 
The UV-optical-FIR SED of the starburst is degenerate, in the sense that
it may be fitted by different set of burst parameters. 
Both models provide a good fit to the UV to sub-mm data.
At radio frequencies however, only the short burst is able to
reproduce the observations while the longer one falls significant
below the observed flux. 
A similar example is shown in Fig.~\ref{arp220} for the ULIRG ARP220.
ARP220 is not consistent with a short burst, though
this could not be excluded by the inspection of UV-optical-FIR data
alone.
Notice that the current SFR of the models differ by about 40\%
and 90\% in M82 and ARP220, respectively.

The cases of M82 and ARP220 show that the inclusion of the radio emission
constitutes a powerful diagnostic tool to investigate
starburst galaxies. Indeed, while the UV, the FIR and the
radio thermal continua are sensitive to the number of living
massive stars, the NT emission is a measure of the current CCSN
rate. Thus the FIR/Radio ratio is a measure of the ratio between
the almost instantaneous SFR and the SFR averaged over the last
few tens of Myrs. Combining the FIR and radio spectral regions is
particularly important for the case of obscured starbursts, where
the burst properties cannot be derived by UV, optical, NIR and
even mid and far IR (continuum) observations.
\begin{figure}
\centering
\includegraphics[angle=270,width=8truecm]{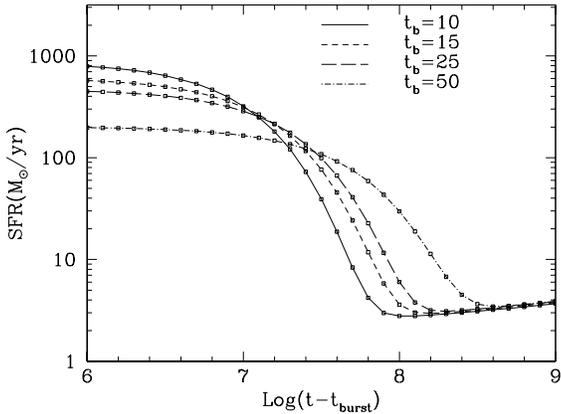}
\caption{The run of the star formation rate in our starburst
models for different e-folding time scales, $t_b$ = 10, 15, 25
and 50 Myr. } \label{sfr}
\end{figure}
We ought to stress in the following, that a careful combination
of FIR and Radio data may be sufficient to determine the recent star
formation history even for those galaxies that are not as
thoroughly observed as M82 or ARP220. Inspection of
Figg.~\ref{m82} and ~\ref{arp220} shows that the expected value of
the q ratio at 1.49 GHz and 8.44 GHz changes by more than a
factor of 2 for the different models. This suggests that
deviations from the average FIR/Radio correlations should be
expected among obscured starbursts and actually could be used as
a powerful diagnostic for the analysis of the star formation in
the burst. Furthermore, also the radio slope is affected
because at younger ages the relative contribution of thermal
emission is larger than at older ages.

To highlight this point we have carefully analysed the evolution
of selected starburst models that possibly encompass different
realistic scenarios. For the star formation rate during the burst
we have assumed four different e-folding time scales, $t_b$=10,
15, 25 and 50 Myr. The mass of stars formed during the burst was
set to 10\% of the underlying disk mass. The obscuration time
t$_{esc}$ was set to linearly decrease with time, from $t_b$ down
to a minimum of 3 Myr, characteristic of normal galaxies. The run
of the SFR of the models during the burst is depicted in Fig.~\ref{sfr}.

\begin{figure}
\centering
\includegraphics[angle=270,width=8truecm]{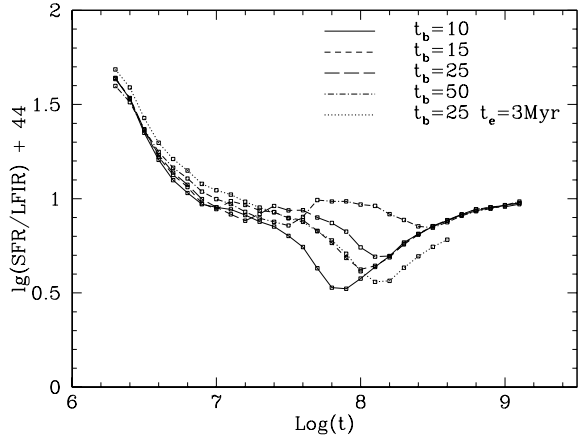}
\caption{Evolution of the ratio between the instantaneous
star formation rate (SFR, M$_\odot$/yr) and  FIR luminosity
(erg/s), see text for the definition.}
\label{clbfir}
\end{figure}

\begin{figure}
\centering
\includegraphics[angle=270,width=8truecm]{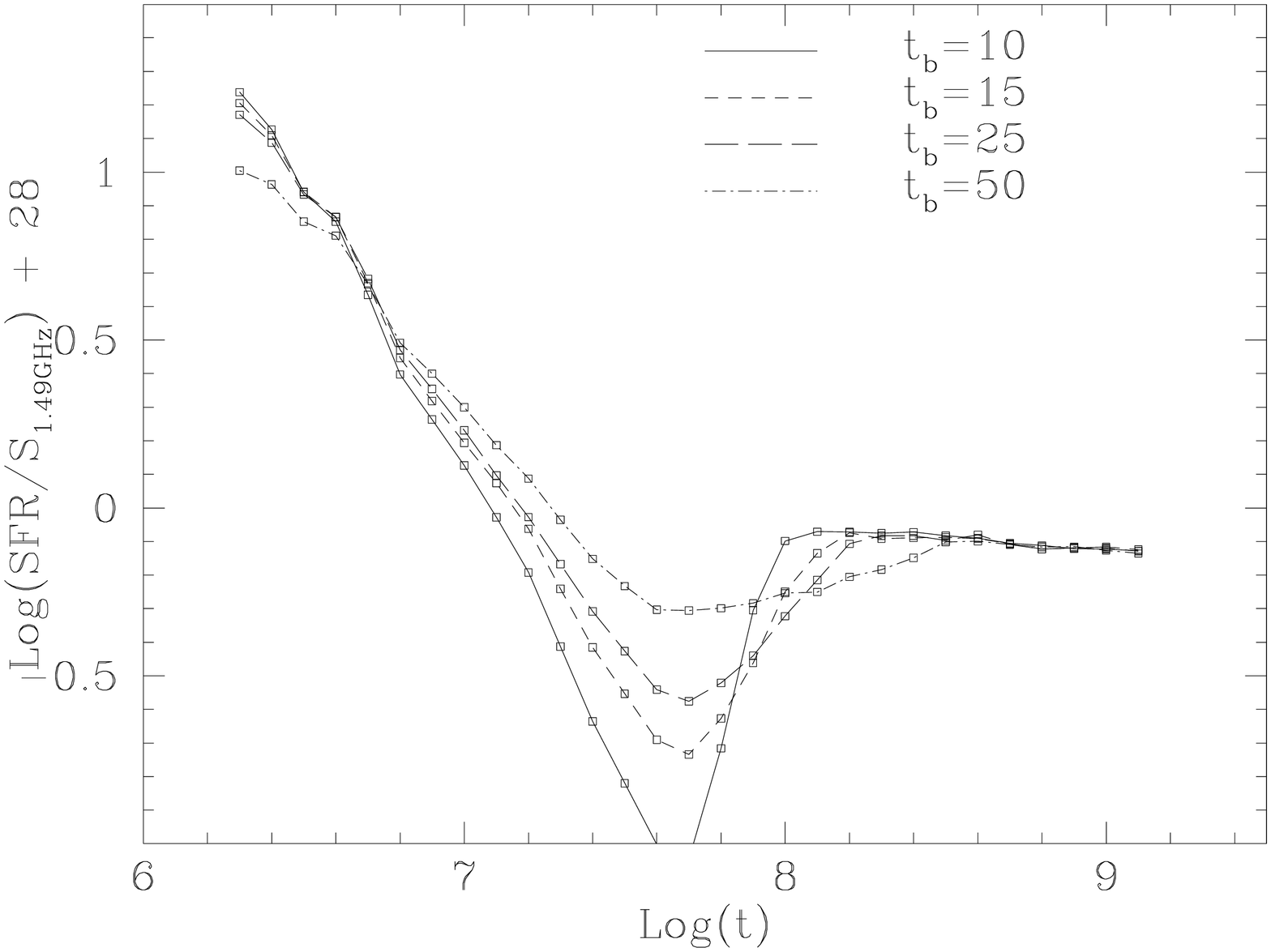}
\includegraphics[angle=270,width=8truecm]{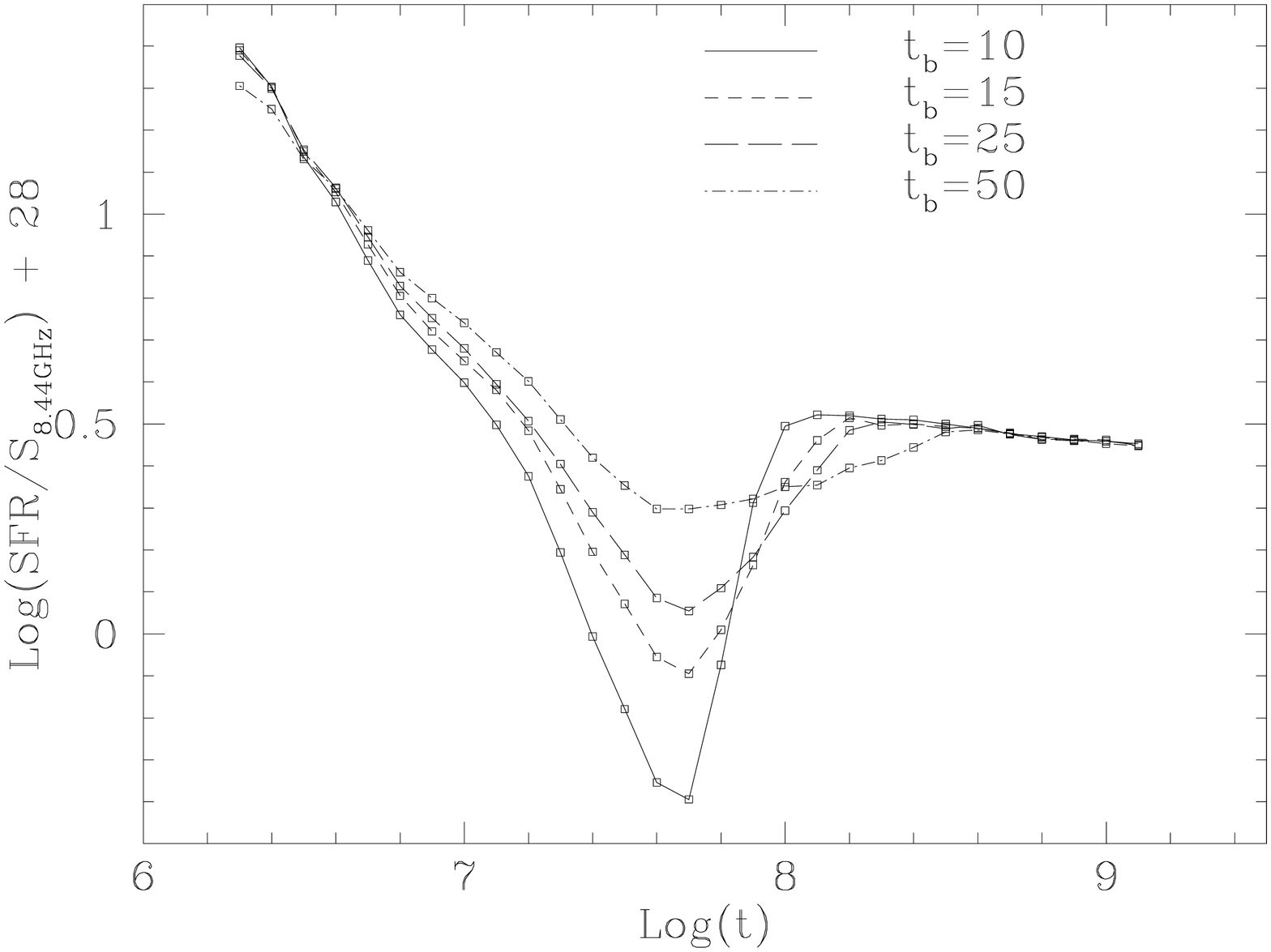}
\caption{Evolution of the ratio between the instantaneous star
formation rate (SFR, M$_\odot$/yr) and radio luminosity (erg/s/Hz)
at 1.49 GHz (upper panel) and at 8.44 GHz (lower panel)}
\label{clbrad}
\end{figure}

Fig.~\ref{clbfir} shows the ratio between the current SFR
(M$_\odot$/yr) and the FIR luminosity, L$_{\rm FIR}$. The latter,
representing a fair measure of the luminosity between 40 and 120
$\mu$m, is defined in analogy with Helou et al.\ (1988) as
L$_{\rm FIR}$(erg/s)$= 1.257 \times 10^{-11} \, (2.58 \, L_{60} +
L_{100})$ where L$_{60}$ and L$_{100}$ are the luminosities in
erg/s/Hz at 60 and 100 $\mu$m. In this figure, the model with
$t_b$=25 Myr has been recomputed with a fixed t$_{esc}$=3 Myr,
corresponding to that of normal galaxies. Fig.~\ref{clbrad} shows
the ratio between the SFR (M$_\odot$/yr) and the radio luminosity
(erg/s/Hz) at 1.49 GHz (upper panel) and at 8.44 GHz (lower
panel). Obviously, in the latter figure, $t_{esc}$
is not relevant.
%(F60 and F100 in Jansky; fIR in erg/cm-2/s)
 Three main evolutionary phases can be recognised, 
the starburst, post-starburst and quiescent phase, characterized by different 
relations between either the FIR or the radio emission and the
current SFR.
At early times, during the starburst phase, the SFR/Flux ratio 
is significantly higher than the asymptotic value defined by normal star forming
galaxies. At later times, when the  ratio falls below the value
of normal galaxies, the models enter the "post-starburst" phase.
Finally, as the burst extinguishes,
the ratio  turns to the asymptotic value defined for normal star forming
galaxies. 
Depending on the
evolutionary status and the e-folding time of the SFR, the
relations between infrared or radio emission and SFR
may change by even one order of magnitude.
As already anticipated, NT radio emission is more
sensitive to the past star formation history than the FIR
emission and the corresponding ratio shows a larger variation.

\section{FIR/Radio properties of compact ULIRGs}
\label{squ}

\begin{table*} 
\caption{Selected galaxies from the sample of compact ULIRGs by Condon et al. (1991)}
\begin{tabular}{lcccccccl}
\hline
    Name        & S$_{1.49GHz}$ &   q & S$_{8.44GHz}$
    & $\alpha_{1.49-8.44}$  &  FIR &N$_e^\#$  & Symbol$^\natural$ &  Classification$^\star$      \\     % I1   &    I0&                  
 IRAS10566+2448     &  46.1 & 2.55&  14.1&  0.68 &      11.90 &-   & H& HII                 \\  %n   -2 &    0 &      IRAS1056+24    
 A11010+4107        &  28.0 & 2.54&  10.7&  0.55 &      11.52 &-   & H& HII                 \\  %n   -2 &    7 &      A1101+41       
 IRAS12112+0305     &  22.6 & 2.66&  10.0&  0.47 &      12.18 &-   & H& HII                 \\  %n   -2 &    8 &      IRAs1211+03    
 UGC08335        &  51.2 & 2.46&  17.0&  0.64 &      11.62    &590 & H& HII                    \\   %590  -2 &    0 &     ARP238         
 IRAS22491-1808     &   6.1 & 3.00&   3.0&  0.41 &      12.02 &-      & H& HII  Tb=5.2 \\%n AGN$<$6\%   -2 &    0 &       IRAS2249-1808   
 IRAS17132+5313  &  28.4 & 2.46&  8.9$^\flat$&  0.67 &     11.79    &450 & H& HII  Tb$<$5            \\ %450  -2 &    0 &     IRAS17132+5313  
 UGC04881        &  29.0 & 2.50&  8.8 &  0.69 &      11.61    &210 & H& HII  Tb$>$7 a          \\   %210  -1 &    0 &     UGC04881       
 IRAS01173+1405  &  43.1 & 2.46&  12.7&   0.70&       11.54   &160 & H& HII  Tb$>$7 a          \\     %160  -1 &    3 &         MCG+02-04-025    
 Mrk331          &  67.5 & 2.51&  21.5$^\flat$&  0.66 &    11.27    &510 & H& HII  Tb$>$7 a       \\    %510  -1 &    0 &            MCG +03-60-036 
 IRAS10173+0828     &   8.8 & 2.92&  5.0  &  0.28 &     11.70 &-   & H&  --             \\   %n    -2    &    0 &     IRAS10173+0828  
 ARP220          & 301.1 & 2.63& 148.0&  0.41 &      12.11    &-   & P& HII  Tb$>$7 a       \\  %n    -1 &    0 &     UGC09913       
\\
 IRAS04191-1855  &  27.3 & 2.49&  8.9 &  0.65 &      11.34    &100 & L& Liner  Tb$<$5      \\   %100   1 &    4 &     ESO 550-IG 025     
 IRAS08572+3915     &   6.5 & 3.11&  4.1 &  0.27 &      11.96 &-   & L& Liner  Tb$<$5      \\%n    1 &    0 &     IRAs0857+39    
 IRAS14348-1447     &  33.2 & 2.38&   9.7&  0.71 &      12.17 &-   & L& Liner  Tb$<$5      \\%n    1 &    11&     IRAS1434-14    
 IRAS01364-1042     &  17.0 & 2.67&  8.2 &  0.42 &      11.67 &-   & L& Liner  Tb$<$5  \\   %n    1 &    0 &      IRAS0136-10    
 UGC08387        & 106.0 & 2.28&  34.9&  0.64 &      11.51    &1220$^\flat$& L& Liner           \\  %1220 ?   1 &    9 & IRAS 13183+3423 
 Mrk273          & 130.0 & 2.31&  43.5&  0.63 &      12.04    &-   & L& Liner  Tb$>$7 a     \\  %n    2 &    10&      UGC08696       
\\
 IRAS03359+1523     &  18.9 & 2.58&  11.0&  0.31 &      11.37 &370 & A& HII  Tb$>$7         \\%370  -1 &    0 &       IRAS0335+15    
 Mrk848          &  46.8 & 2.38&  12.1$^\flat$&  0.78 &    11.72    &240 & A& HII  Tb$>$7      \\   %240   4 &    0 &     IZw107             
 UGC02369        &  42.7 & 2.39&  13.3&  0.67 &      11.42    &-   & A& HII Tb$>$7 l   \\   %n    5 &    2 &      UGC02369       
 IIIZw035        &  39.3 & 2.58&  19.7&  0.40 &      11.46    &590 & A& Liner Tb=5.4 \\ %590   1 &    1 &     IIIZw035       
 IRAS15250+3608     &  12.8 & 2.81&  10.5&  0.11 &      11.88 &-   & A& Liner  Tb=6       \\    %n    4 &    12&      IRAS1525+36    
 UGC0bright        & 146.0 & 2.10&  52.6&  0.59 &      11.93    &-   &A & Liner Tb$>$7 l \\ % 5  UGC05101       6   Liner-Tb>7L     
 NGC2623         &  97.8 & 2.51&  35.5&  0.58 &      11.47    &-   & A& AGN  Tb$>$7 l       \\  %n    5 &    5 &      NGC2623        
\\
 NGC0034         &  58.7 & 2.52&  15.2&  0.78 &      11.28    &-   & S& Sey2  Tb=5          \\  %n    3 &    0 &      MRK938         
 Zw475.056       &  26.0 & 2.65&   8.2&  0.67 &      11.37    &190 & S& Sey2  Tb$<$5         \\ %190    3 &    13&        Zw475.056      
 IRAS05189-2524     &  28.1 & 2.76&  11.4&  0.52 &      11.91 &-   & S& Sey2  Tb$<$5       \\%n    4 &    0 &     IRAS0518-25    
 Mrk231          & 240.0 & 2.24& 265.0& -0.06 &      12.35    &-   & M& Sey1  Tb$>$7        \\  %n    5 &    0 &      UGC08058       
\hline
\end{tabular}

{\footnotesize
$^\flat$ Uncertain value\\
$^\#$ Electron density, cm$^{-3}$, from Veilleux et al. (1995)\\
$^\natural$ Symbols adopted in Figg \ref{figfir} and \ref{allslo}\\
$^\star$ HII, Liner, AGN, Seyfert based on Veilleux et al. (1995); 
$T_b$ logarithm of the brightness
temperature from VLBI data by Smith, Lonsdale \& Lonsdale (1998)\\}
\label{ulirgs}

\end{table*}

We have already anticipated that our models predict the existence of a correlation
between the radio slope and the FIR/radio ratio, because both quantities are
affected by the relative contributions of FIR, thermal and non-thermal radio
emission that change during the evolution of the burst. Understanding whether the
use of such correlation can provide further physical insight on the star formation
process becomes, obviously, particularly relevant in the case of highly obscured
starbursts such as ULIRGs, emitting the bulk of their radiation in the mid and  far
infrared. ULIRGs are characterized by extreme FIR luminosities exceeding the UV-
optical power by even more than one order of magnitude, and are believed to be
transient phases of galaxy activity associated with the dynamical interaction and
merging of gas rich systems. Whether their powerful emission is of starburst
origin or is due to the AGN is still under debate (Sanders et al. 1988, Condon et
al 1991, Veilleux et al 1994) but, recently, it has become clear that these objects
may harbour huge compact star forming complexes and the AGN, at the same time
(Rowan-Robinson, 1995). However, it remains still unclear what fraction of the
bolometric luminosity is eventually provided by the central monster and at what
level this affects FIR and radio emission. For example, very recently Berta et al.
(2002) and have shown that in IRAS 19254--7245 ("The Superantennae"), the AGN is
contributing about 40--50\% of the MIR and FIR emission ( essentially the
bolometric luminosity), while in the radio it overcomes by one order of magnitude
the emission from the starburst. At the same time, for IRAS 20100--4156, one of the
brightest nearby objects, both the multi-wavelegth analysis (Fritz et al 2002),
spectro--polarimetry (Pernechele et al 2002) and near-IR spectroscopy (FWHM
Pa$\alpha\simeq$440Km~s$^{-1} $, Valdes et al. 2002) show no evidence of the AGN.

In this respect, the sample of compact ULIRGs selected by Condon et al. (1991)
turns out to be particularly interesting because high resolution VLA maps suggest
that the bulk of their radio continuum could be of 
starburst origin and several objects have been
observed at both 1.49GHz and 8.44GHz, so that a direct comparison in term of
"observables" can be made with our starburst models.
We thus isolated all the compact ULIRGs of the Condon et al. sample
observed at both 1.49GHz and 8.44GHz, in Table \ref{ulirgs}.
Optical spectroscopic classification by Veilleux et al (1995) indeed showed that a
number of the selected objects have emission line ratio characteristic of
photoionization by massive stars (HII galaxies), but other show higher excitation
and were classified as Liners or even Seyfert 2 galaxies. Mrk~231 is a
Seyfert 1 object. A subset of the original Condon et al. sample has been
subsequently mapped with VLBI by Smith, Lonsdale \& Lonsdale (1998, SLL), to establish the
nature of the milli--arc seconds structures and in particular to obtain the
brightness temperature of the compact inner core.
SLL did not found significant correlations between the 
VLBI emission (usually 10\% of the 1.6GHz total flux density)
and other physical parameters like  total radio power, FIR emission and radio slope.
Perhaps more important, they have found that
that optical excitation is not strictly correlated with high T$_b$.
As can be seen from their data, summarized together with the Veilleux et al.
classification, in the last column of Table \ref{ulirgs}, there are
HII objects showing high T$_b$ cores and LINERs and Sey 2 (all) galaxies
with central T$_b$ consistent with
a starburst origin. Furthermore, SLL were able to show that
some of the high T$_b$ cores could be explained by 
bright radio supernovae complexes, consistent with the star formation rates
required by the FIR. These objects are marked with a lower case "a"
in the last column of Table \ref{ulirgs}, while those that could
not be interpreted in terms of bright SNRs by SLL,
are marked by a lower case "l". Some of the
high T$_b$ objects had not enough detailed structure to allow this 
analysis.

According to  the information provided by Veilleux et al. (1995) and 
SLL, we grouped all objects into four broad
categories, named with the symbol reported in column 8 of Table \ref{ulirgs}. The
letter "H" indicates all galaxies of HII type, but for three objects for which SLL
could not establish that their high $T_b$ is compatible with RSN
complexes. Analogously, Liners are indicated by the symbol "L" when their $T_b$ is
missing or is low, or their high $T_b$ was found compatible with RSN complexes. Object
classified as HII, Liners or AGN that do not fulfil the above requirements are
indicated by the symbol "A" while, for Seyfert 2 galaxies we use the symbol "S".
Finally "P" and "M" refer to ARP220, a typical starburst galaxy and Mrk~231, a
Seyfert 1 galaxy, respectively. According to this grouping, the effect of the AGN,
if any, should not be significant in types H to L. For example a recent
comprehensive analysis of optical-NIR-FIR SED and of spectro-polarimetric data of
IRAS 22491-1808, indicates that if the AGN is present in this object, it must
contribute less than 5\% of the total FIR emission (Fritz et al 2002).

\begin{figure}
\centering
\includegraphics[angle=270,width=8truecm]{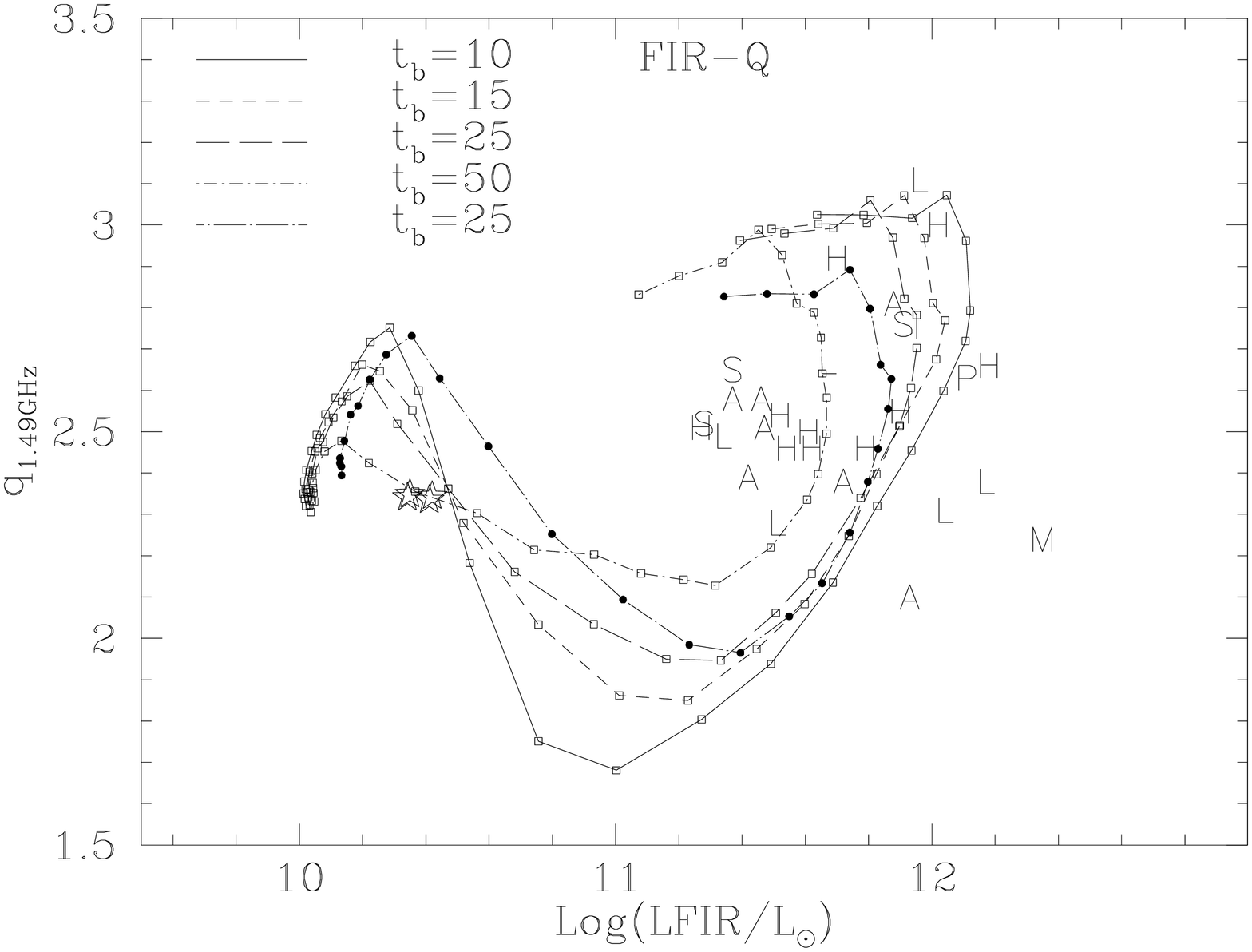}
\includegraphics[angle=270,width=8truecm]{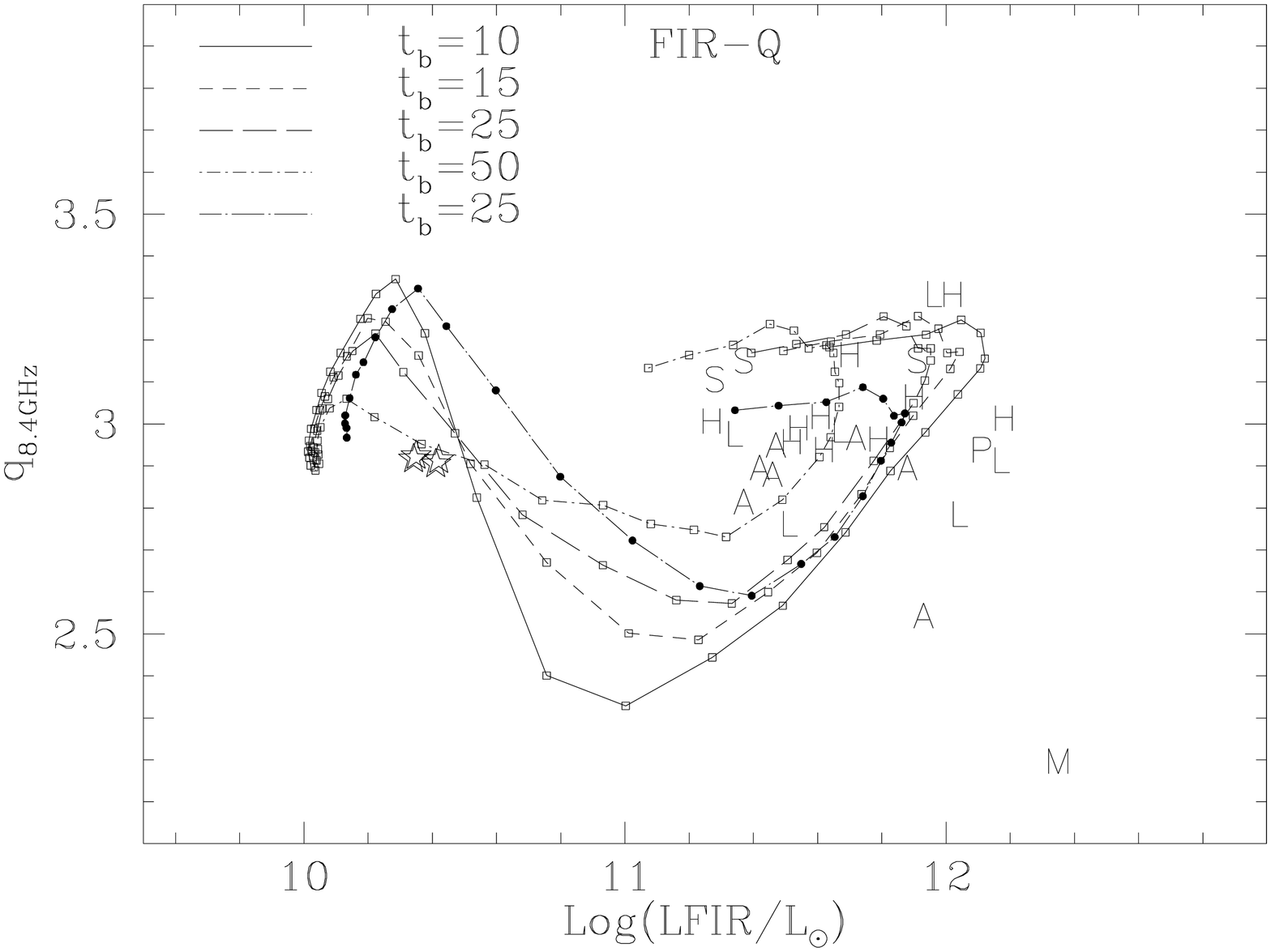}
\caption{Evolution of starburst models in the q vs L$_{\rm FIR}$ diagram. 
The upper panel is for q at 1.49 GHz, the lower one
is for q at 8.44 GHz. Stars indicate the positions of
typical models of quiescent star forming disks. The model with
$t_b$=25 Myr and t$_{esc}$=3 Myr is also shown (small solid
dots). Symbols refer to the subset of compact ULIRGs (Condon et al.
1991), drawn in Table \ref{ulirgs}. } \label{figfir}
\end{figure}

In Figure~\ref{figfir} we show the run of the q values expected
from our starburst models, Eq.(\ref{qeq}), at 1.49 and 8.44
GHz, against the FIR luminosity (now in solar units).
Superimposed are  the selected compact ULIRGs from the local
sample of Condon et al. (1991), with the corresponding symbols
listed in Table \ref{ulirgs}. 

The models can be arbitrarily
shifted in the horizontal direction by changing the mass of stars
formed in the burst which, for the adopted normalization,
corresponds to about 10$^{10}$ M$_\odot$. To fit the IR
luminosity of the brightest compact sources we need a SFR higher
by about a factor of 2 to 3, for the shorter burst. We note that
the observed dispersion of the q ratio is well explained by the starburst
models, in particular in the lower panel, where, as we will see,
the effects of free-free absorption should be negligible.
%\begin{figure}[]
%\psfig{file=etavarclb.ps,width=8cm}
%\caption{Evolution of the ratio between the current (t=0)
%star formation rate and  FIR emission (upper panel)
%or  radio emission (lower panel)}
%\label{allclb}
%\end{figure}
The mass of the burst is however not known a priori and while
Fig.~\ref{figfir} shows  that the observed data are consistent
with the sources being compact starbursts, it does not allow to
obtain an estimate of the burst parameters. On the other hand,
the case of M82 and ARP220 (Figg.~\ref{m82} and \ref{arp220})
indicates that this is within the possibility of the model when
enough observations are available.

\begin{figure}
\centering
\includegraphics[angle=270,width=8truecm]{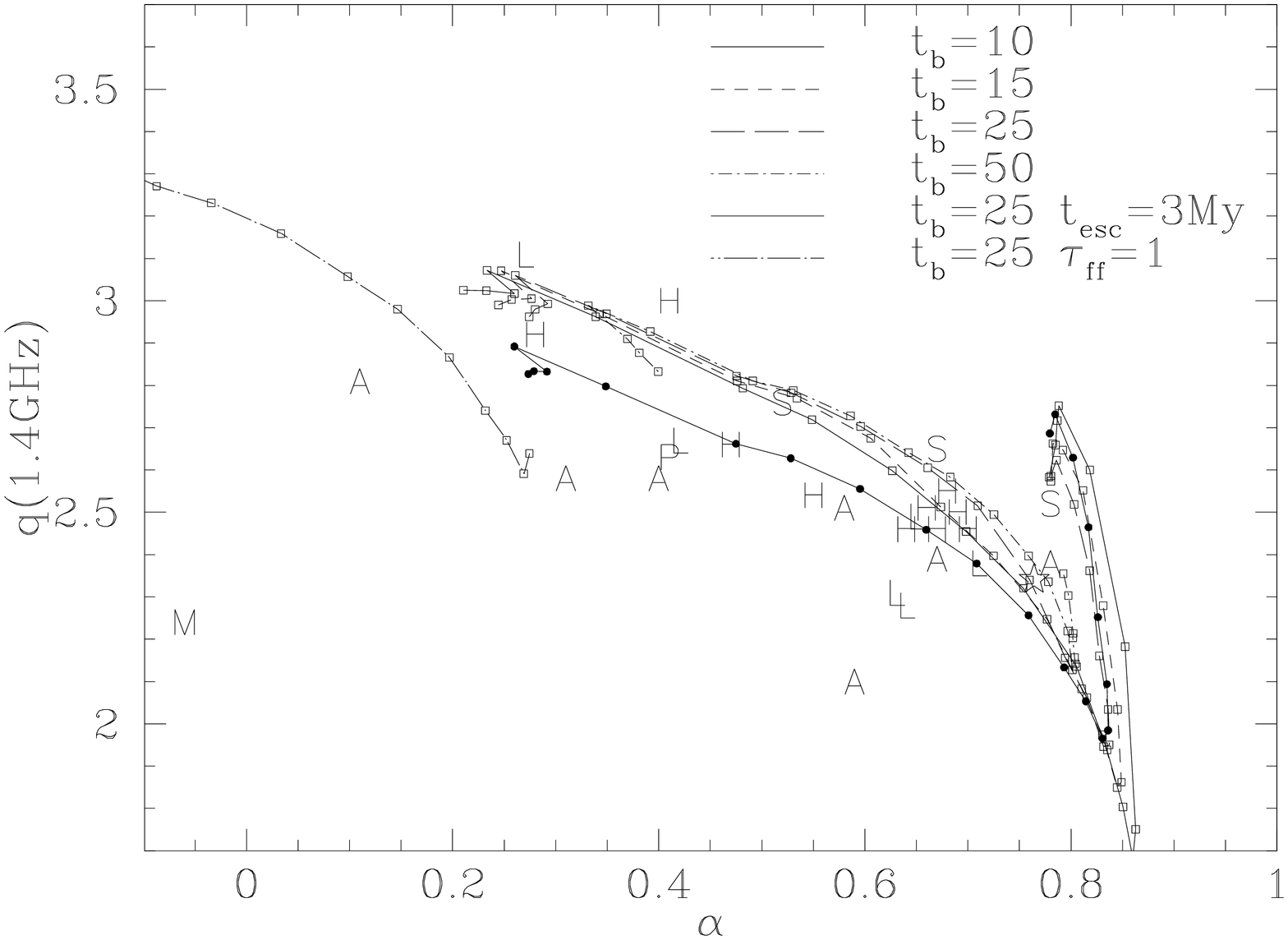}
\includegraphics[angle=270,width=8truecm]{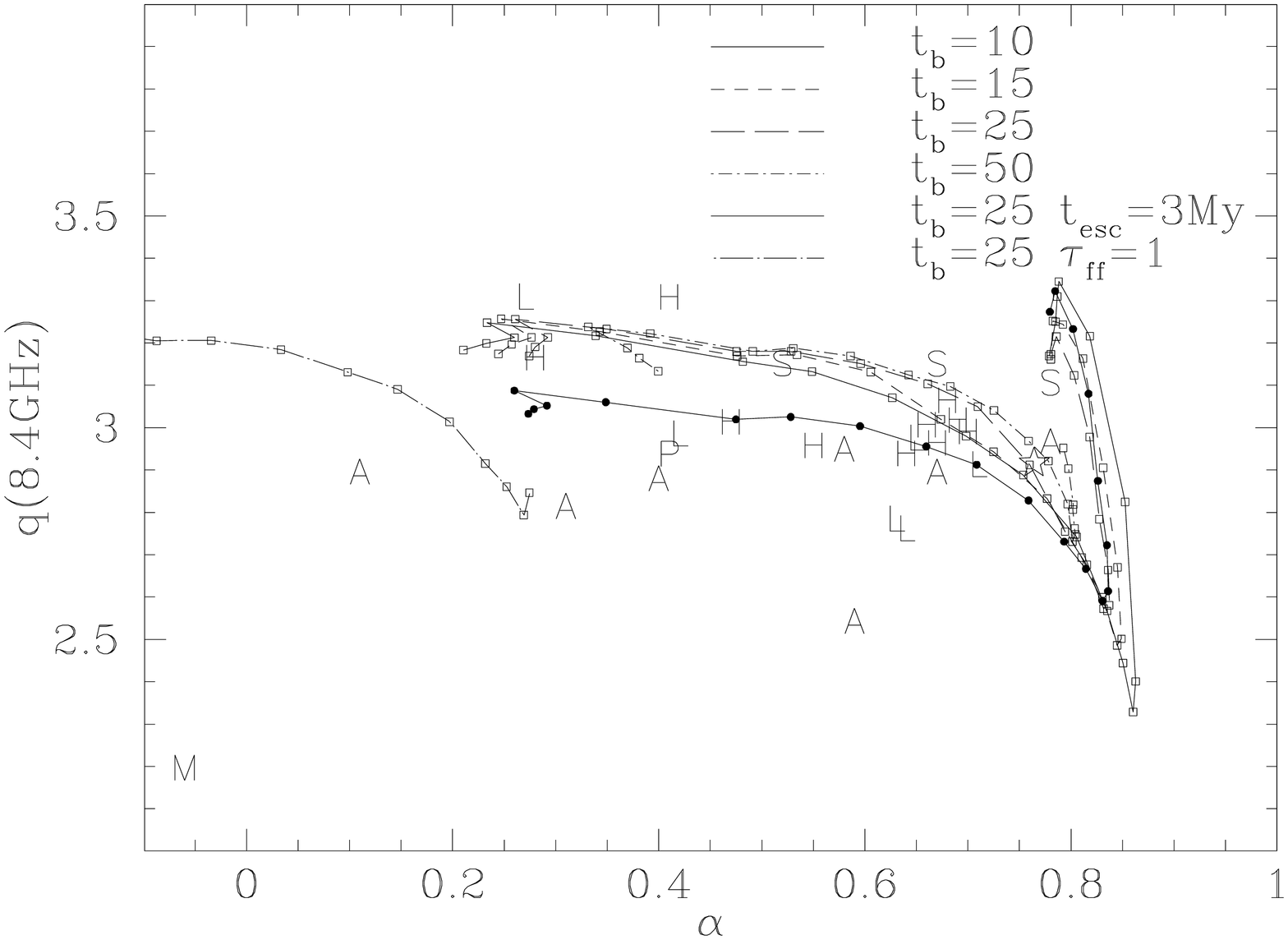}
\caption{Upper panel: evolution of the ratio q vs radio slope
$\alpha$ between 8.4 and 1.4
GHz for our  starburst models. Characteristic
starburst ages range from log $t$ (yr) = 6.3 to 8.6. The model
with $t_b$=25 Myr has been recomputed both with a t$_{esc}$=3 Myr
(crossing the observed data) and by adopting a free-free
absorption with $\tau_{\nu=1.4GHz} = 1$ (model on the left side
up to its minimum value of q). Lower panel: same for $\nu$=8.4
GHz. Note that q is no more affected by free-free absorption.
Symbols refer
to the subset of compact ULIRGs (Condon et al.
1991), drawn in Table \ref{ulirgs}. 
}
\label{allslo}
\end{figure}

We thus combine in Fig.~\ref{allslo} the FIR/Radio ratio with the
slope of the radio emission $\alpha$ between 1.49 and 8.4 Ghz.
Both quantities are independent from the intensity of the burst.
Rather, they depend on the form of the recent star formation
history, so that the path in this diagram traces the evolutionary
status of the starburst. The radio slope of the models changes
because of the variation of the dominant source of radio emission
as the starburst ages. During the first 3 to 4 Myrs only thermal
emission from HII regions contributes to the radio emission, and
the radio flux has a characteristic slope $\alpha\simeq$0.1. Then
CCSN explosions feed relativistic electrons into the galactic
magnetic field, and NT emission steepens the spectrum toward a
slope which is more typical of normal galaxies. At the same time
the total radio power increases and the evolution of the q ratio,
though affected by the corresponding increase of the FIR
emission, continuously decrease to a minimum value. At this stage
the SFR has decreased significantly  and the model can be
considered in a post starburst phase. At even older ages the
ratio increases again, but this corresponds to the very late
phase where the model can no longer be considered representative
of an ultra luminous galaxy.
The model with $t_b$=25 Myr and obscuration time of normal
galaxies, t$_{esc}$=3 Myr, is not able to reproduce the observed
high values of q, in both panels, confirming that the low UV flux
observed in other obscured starburst (see e.g.\ Figg.~\ref{m82}
and \ref{arp220}) indicates that this escape time is
too short. The observed data show a clear trend of increasing q
at decreasing slope, in the 1.49 GHz plot, with the models being
only able to delineate an upper envelope. At a higher frequency
(lower panel), the trend in q disappears while the models still
delineate the upper envelope. A possible explanation is that part
of the effect is caused by free-free absorption (see also Condon
et al.\ 1991). At radio frequencies the optical depth of
free-free absorption by a cloud with electron density $N_e$
and size l, is
\begin{equation}
\tau_\nu^{ff} \simeq 8.2\times{10^{-2}} \,
T_e^{-1.35}(\frac{\nu}{\mbox{GHz}})^{-2.1}(\frac{N_e^2\,l}{\mbox{pc/cm}^6})
\label{tauff}
\end{equation}
In order to highlight the effects of free-free absorption we have
recomputed a model with optical depth $\tau^{ff}_{1.49{\rm
Ghz}}$=1 (dot-dashed line on the left side in Fig.~\ref{allslo}).
Assuming an electron temperature T$_e=10^4$ K, this optical depth
corresponds to an emission measure of about 6$\times$10$^6$
pc/cm$^6$. The latter can be achieved by considering
2.5$\times$10$^8$M$_\odot$  within 250 pc or
7.8$\times$10$^8$M$_\odot$ within 390 pc of ionised gas, for an
average electron density of about $N_e$=150 cm$^{-3}$ or
$N_e$=125 cm$^{-3}$ and solar composition, respectively.  The
above figures compare fairly well with electron densities 
derived from emission lines ratio by Veilleux et al. (1995), listed in Table
\ref{ulirgs}.
They are consistent both with the fact that the selected
ULIRGs are compact radio sources with typical sizes less than 1
Kpc and possibly as low as a few hundred pc (Soifer et al.\
2000), and that the total mass involved in our starburst models
is of the order of 10$^{10}$M$_\odot$.

The distribution of the data in Fig.~\ref{allslo} also renders
quite unlikely the possibility that the slope variation is due to
an increasing importance of the electron cooling by inverse
Compton. Indeed it could be that in ULIRGs with the lower
$\alpha$ electrons cool down  by inverse Compton on the stellar
radiation field: in those objects only thermal emission would be
present at radio wavelengths. Not only this would again require a
fine tuning of the two cooling processes, in order to give rise
to a distribution of slopes between 0.1 and 1 but, and more
important, it would be difficult to explain the absence of the
trend at the higher frequency. This enforces our interpretation
in terms of free-free absorption because, given the frequency
dependence (Eq.~\ref{tauff}), this effect should not affect the
value of q at 8.44 GHz. The quantity q$_{8.44 \mbox{GHz}}$, which
still shows a range of about 0.6dexp, can be considered as a
genuine measure of the age of the compact starbursts.
Unfortunately, in this diagram, the slope is still affected by
free-free absorption, and it is not possible from the above data
alone to identify precisely the evolutionary status of the
ULIRGs. However,  our models suggest that, observing ultra
luminous galaxies at frequencies between 8.44 GHz and 23 GHz
would possibly constitute a powerful tool to investigate on the
recent star formation history of obscured starbursts.

From the upper panel of Fig.~\ref{allslo} it is also evident that
there is no sharp threshold value of q$_{1.49\mbox{GHz}}$ that
can be safely used to delineate a separation between star
formation and AGN powered ULIRGs. 
 There is a tendency for objects classified "A" 
to occupy the lower boundary allowed by starburst models
but, even in the case of the
Seyfert 1 galaxy Mrk~231,
the FIR brightest object in our sample,
it would be difficult to exclude a
starburst origin resting only on its q$_{1.49\mbox{GHz}}$.
However its location in the corresponding q-slope diagram is not
matched by any of our starburst models and, by looking at the
higher frequency data q$_{8.44\mbox{GHz}}$, it appears to be at
least three times more radio powerful (relative to the FIR) than
all the other sources and than allowed by our models in the
starburst phase. Is this an indication that the parameters plotted
in Fig.~\ref{allslo} (possibly making use of a higher frequency to
avoid free-free contamination of the radio slope) could be used as
a diagnostic diagram to disentangle a starburst from an AGN? 
And, in this case, why are all the Seyfert 2 galaxies in our sample
just populating the upper envelope of starburst models?. From
the above plot we may only conclude that a minimum
value of q$_{1.49\mbox{GHz}}$ for a starburst powered source is
around 1.8, and this may set a threshold value below which
another source of radio emission has to be invoked. One should
consider however that at this frequency the nuclear engine may be
masked by free-free absorption. At 8.44 GHz free-free absorption
should be less important and a fair threshold for star formation
dominated ULIRGs is q$_{8.44\mbox{GHz}}$=2.5. 

\begin{figure}
\centering
\includegraphics[angle=270,width=8truecm]{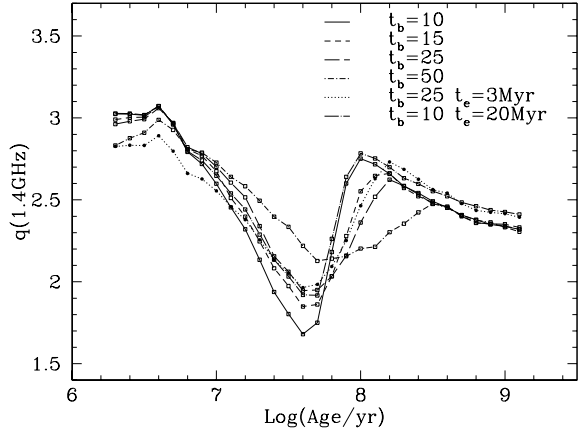}
\caption{The q ratio as a function of time for the models in
Fig.~\ref{allslo}.} \label{allqt}
\end{figure}

%%%%%%%%%%%%%%%%%%%%%%%%%%%%%%%%%%

\section{The post starburst phase}
\label{spsbt}

In a recent comprehensive analysis of the FIR/Radio correlation
in nearby Abell clusters, Miller \& Owen (2001) found a
statistically significant excess of star forming galaxies with
enhanced radio emission relative to the FIR, toward the cluster
centres. High resolution radio images have also excluded a
significant AGN contribution to the radio emission in these
galaxies.

There is a long standing debate on the nature of this excess of
galaxies with a low value of q$_{1.49\mbox{GHz}}$, which is
always interpreted as a radio enhancement and may reach a factor
of three. Gavazzi \& Jaffe (1986) advanced the hypothesis that
the radio excess is caused by the ram pressure strengthening of
the galaxy magnetic field, as the galaxy travels through the
intracluster medium. According to Miller \& Owen (2001) this
explanation seems not to work because of the lack of any
correlation with the galaxy velocity. They suggest instead a
compression of the galactic magnetic field by thermal pressure of
the intracluster medium.

However, due to its small scatter, the FIR/Radio correlation
should be quite independent from the effects of the environment.
We thus advance the hypothesis that the observed excess is due to
an excess of post starburst galaxies in the central regions of the
cluster. Indeed soon after a peaked starburst episode and/or a
star formation interruption, the radio emission fades less rapidly
than the FIR emission, causing an apparent radio enhancement. In
this case the effect would be only indirectly due to the
environment, as the gas rich galaxies enhance and/or exhaust
their star formation through a central crossing.

All our models show a more or less pronounced minimum value of q
during the post starburst phase. The model with $t_b$=10 Myr
remains below q$_{1.49{\rm GHz}}$=2 for about 20 Myr (Fig.~\ref{allqt}),
while its luminosity at the minimum q has decreased by one order
of magnitude (Fig.~\ref{figfir}), but it is still infrared
luminous. By decreasing the total mass formed in the burst it is
easy to populate for a short time the region with q$_{1.49{\rm GHz}}$
below 2 and L$_{\rm FIR}$ between 10$^{10}$ and
10$^{11}$ L$_\odot$. Thus a low value of q$_{1.4 {\rm GHz}}$ could simply
be a natural consequence of the particular star formation history
experienced by the galaxy. Because high resolution radio images
probe that the emission is extended, the recent star formation
history, either simply interrupted or enhanced and exhausted in a
burst, must have been globally synchronised. Our models place an
upper limit of less than 100 Myr to the age of the last major
burst and/or interruption of the star formation. At a typical
velocity of 1000 Km/s, the galaxy has moved by only 100 Kpc since
the beginning of the burst. This figure could be about two times
larger if we allow for the formation of CCSN down to 5 M$_\odot$
(overshoot models without O, Ne, Mg white dwarfs) and use a
slightly slower star formation decline.

In summary, our models indicate that the star formation
switch-off happened not too much far from the present galaxy
position. The excess of low q galaxies is thus simply due to the
larger probability of switching off the star formation in the
higher density regions of the cluster. It is a nurture effect,
possibly due to the higher degree of harassment suffered in the
central regions. As a clear implication, our spectrophotometric
models predict that these galaxies should show enhanced Balmer
absorption features, e.g.\ EW H$\delta\geq$5\AA (see also Miller
\& Owen 2001).
If our interpretation is correct, the analysis of the statistics
of the deviation from the FIR/Radio correlation at values of q
lower than the average, coupled with the low characteristic time
involved, provides an independent measure of the local rate at
which the Butcher-Oemler effect is operating within galaxy
clusters.

\section{High redshift starbursts}
\label{shz}

In the last few years, a wealth of observations performed in FIR/sub-mm spectral
regions revealed the existence of a new class of galaxies interpreted as the high-
z analogue of the local ULIRGs (e.g.\ Smail, Ivison \& Blain 1997, Hughes et al.\
1998, Barger, Cowie, \& Sanders 1999). The sub-mm fluxes, probably mostly powered
by star formation rather than AGN, (Granato, Danese, \& Franceschini, 1997;
Almaini, Lawrence, \& Boyle, 1999), imply star formation rates of several hundreds
of M$_\odot$/yr.

The discovery of these galaxies has introduced a new test to the
theories of structure formation, but to this aim the knowledge of
their redshift distribution is of fundamental importance. Due to
the uncertain position, or to the lack, of the optical
counterparts, a spectroscopic redshift is available only for a few
sub-mm sources. Instead, an estimate of the redshift for most of
these galaxies has been performed exploiting the FIR-radio
correlation observed for local star forming galaxies, under the
hypothesis that high-z galaxies obey the same correlation as
local ones (Carilli \& Yun 1999, 2000; Dunne, Clements, \& Eales
2000; Yun \& Carilli 2002).

Due to the very different power laws of the sub-mm and the radio spectra, the
spectral index $s^{353}_{1.4} \equiv \log(S_{353}/S_{1.4})/\log(353/1.4)$, between
the 850 $\mu$m ($353$ GHz) and the $1.4$ GHz fluxes, depends very strongly on
redshift, with the main uncertain factors in its effectiveness as a redshift
indicator being (Carilli \& Yun 2000; Dunne et al.\ 2000): (a) the dust
temperature distribution, (b) the slope of the dust emissivity in the sub-mm and
(c) the slope of the radio spectrum. Additional uncertainties are
(d) the assumed validity of the FIR-radio correlation at high-z, that recently has
been directly tested to z$\sim 1.3$ (Garrett 2002), (e) the possible presence of
AGNs (whose values of $s^{353}_{1.4}$ could be attributed to lower z star forming
galaxies rather than high-z AGN), and (f) the quenching of the synchrotron photons
due to inverse Compton of relativistic electrons off the CMB radiation, which is
expected to be important for $z > 6$ (Carilli \& Yun 1999).

With our model, we have investigated the $s^{353}_{1.4}(z)$
relation for starburst galaxies, keeping into account the
uncertainties a, b and c. These factors affect the
intrinsic shape of the galaxy SED. Uncertainties due also to observational and calibration errors are
accounted for by Hughes et al.\ (2001).
The T distribution of dust and its
slope are degenerate in the sub-mm, because a shallower slope can
mimic the effect of a colder dust component with a steeper slope
(e.g.\ Silva 1999). This is due to the fact that the convolution
of gray bodies of different temperatures (resulting from the T
distribution of dust within a galaxy) yields a slope shallower
than the effective one of the emissivity of dust. Moreover both
the T distribution of dust in galaxies and the slope of the radio
emission depend on several factors that change during the
evolution of the starburst (i.e.\ the evolution of the stars
heating the dust and their distribution across the galaxy, the
dust optical depth, the relative importance of free-free and
synchrotron emission and the amount of free-free absorption, see
Sect.~\ref{sq}).

\begin{figure}
\begin{center}
\includegraphics[width=8truecm]{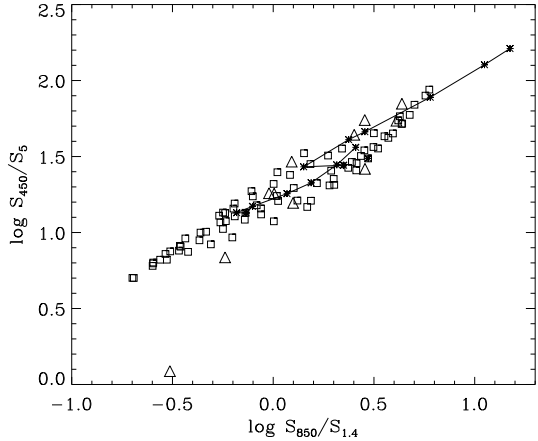}
\caption{Color--color plot $\log S_{850 \mu m}/S_{1.4 GHz}$ vs $\log S_{450 \mu m}/S_{5 GHz}$.
{\it Squares}: Starburst models with $t_b=5, 10, 15, 25, 50$ Myr evolved to an age of $3 \, t_b$.
{\it Diamonds}: models for M82 and ARP220 (see Figg.~\ref{m82} and \ref{arp220}), reproducing the SED from the UV to the radio
(see Silva et al.\ 1998 for details).
{\it Asteriscs} connected by solid line: evolution during the starburst of the models reproducing
M82 and ARP220.
{\it Triangles}: Data from Dunne et al.\ (2000), Condon et al.\ (1991).}
\label{colcol}
\end{center}
\end{figure}

Therefore we have adopted the set of starburst models described
in Sect.~\ref{sq}, i.e. the exponentially decreasing starbursts
with e--folding times $t_b=10$, 15, 25, 50 Myr, complemented with
one more case, $t_b=5$ Myr, and with the models for M82 and ARP220
(Figg. \ref{m82} and \ref{arp220}), to study the dependence of
$s^{353}_{1.4}(z)$ on the evolutionary status of the starbursts
and the consequent different shapes of the SEDs. In particular,
the model for ARP220 needs a slope for the emissivity of dust in
the sub-mm of 1.5, while this is not required for M82 and other
nearby galaxies (see Silva et al. 1998), whose SEDs can be
reproduced with a slope of 2. In Sect.~\ref{sq} we showed that
the starburst models cover the range of observed values of the
FIR-radio relation and of the radio spectral index. In
Fig.~\ref{colcol} we have checked that the range of values of the
sub-mm to radio ratios ($850 \mu$m$-1.4$GHz and $450 \mu$m$-5$GHz
to check also the spectral regions observed at high z) of the
models, cover at least all the observed range. Furthermore, we
can take into account also ratios not observed in the available
local galaxy samples, but that might be expected to be present in
some phases (e.g.\ during the first $\sim 3$ Myr since the onset
of the starburst when the radio emission is mostly thermal). The
models are evolved to an age of $3\,t_b$ since the start of the
burst.
\begin{figure}
\begin{center}
\includegraphics[width=8truecm]{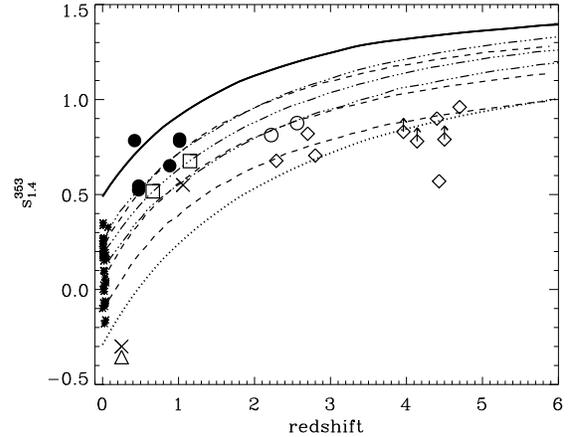}
\caption{The spectral index $s^{353}_{1.4}$ as a function of redshit
for our models (upper continuos and lower dotted lines), and from Carilli \& Yun (2000)
and Dunne et al.\ (2000) (respectively dashed and 3 dot--dashed lines for average and $\pm 1 \sigma$).
The upper model curve corresponds to a very initial phase of the starburst, when free free radio emission dominates, while
the lowest one is for a late phase.
Data for galaxies and AGN with spectroscopic redshift are reported:
{\it Asteriscs} are local galaxies by Carilli \& Yun (2000) and Dunne et al.\ (2000);
{\it Diamonds} are AGN by Rowan-Robinson et al.\ (1993), Ivison et al.\ (1998), Yun et al.\ (2000);
{\it Filled circles}, {\it Crosses}, {\it Squares}, and {\it Circles} are sub-mm and radio selected galaxies
from Barger et al.\ (2000), Smail et al.\ (2000), Eales et al.\ (2000), and Ivison et al.\ (2000)
respectively. The {\it Triangle} is a cD galaxy from Edge et al.\ (1999).}
%(le curve sono arp220 dopo $1e6$ anni dall inizio del burst e
% $t_b$=10, log age = $7.5$)
\label{alfa_z35314}
\end{center}
\end{figure}
In Fig.~\ref{alfa_z35314} we show the $s^{353}_{1.4}$ vs z
relation resulting by considering all our models (we plot the
highest and the lowest one, corresponding respectively to a very
initial phase dominated by free free thermal radio emission and a
late phase, dominated by NT radio emission), together with the
relations by Carilli \& Yun (2000) and Dunne et al.\ (2000). As
already remarked by these authors, the effect of low frequency
free-free absorption on the relation is important only at very
low z, due to its strong frequency dependence ($\tau_\nu^{ff}
\propto \nu^{-2.1}$).

We note that the dispersion of local observations is entirely
compatible with the one introduced by the evolution of the
starburst. Furthermore, the different evolutionary stages and
thus SED shapes of the models, result in a significant dispersion
of $s^{353}_{1.4}$ at any z, sufficient to invalidate any
quantitative estimate of z. We show that it might be possible to
reduce the uncertainty in the redshift estimate by combining
$s^{353}_{1.4}$ with another spectral index that depends mainly
on the starburst age. The latter is a radio spectral index (see
Sect.~\ref{sq} and Fig.~\ref{allslo}). Indeed for redshift
between 0 and 6 the radio spectral index of each model changes
less than 0.2. Thus one may use this index to
confine the $s^{353}_{1.4}$-z evolution within the possible
values allowed by the corresponding selected starburst phase.

\begin{figure}
\begin{center}
\includegraphics[width=8truecm]{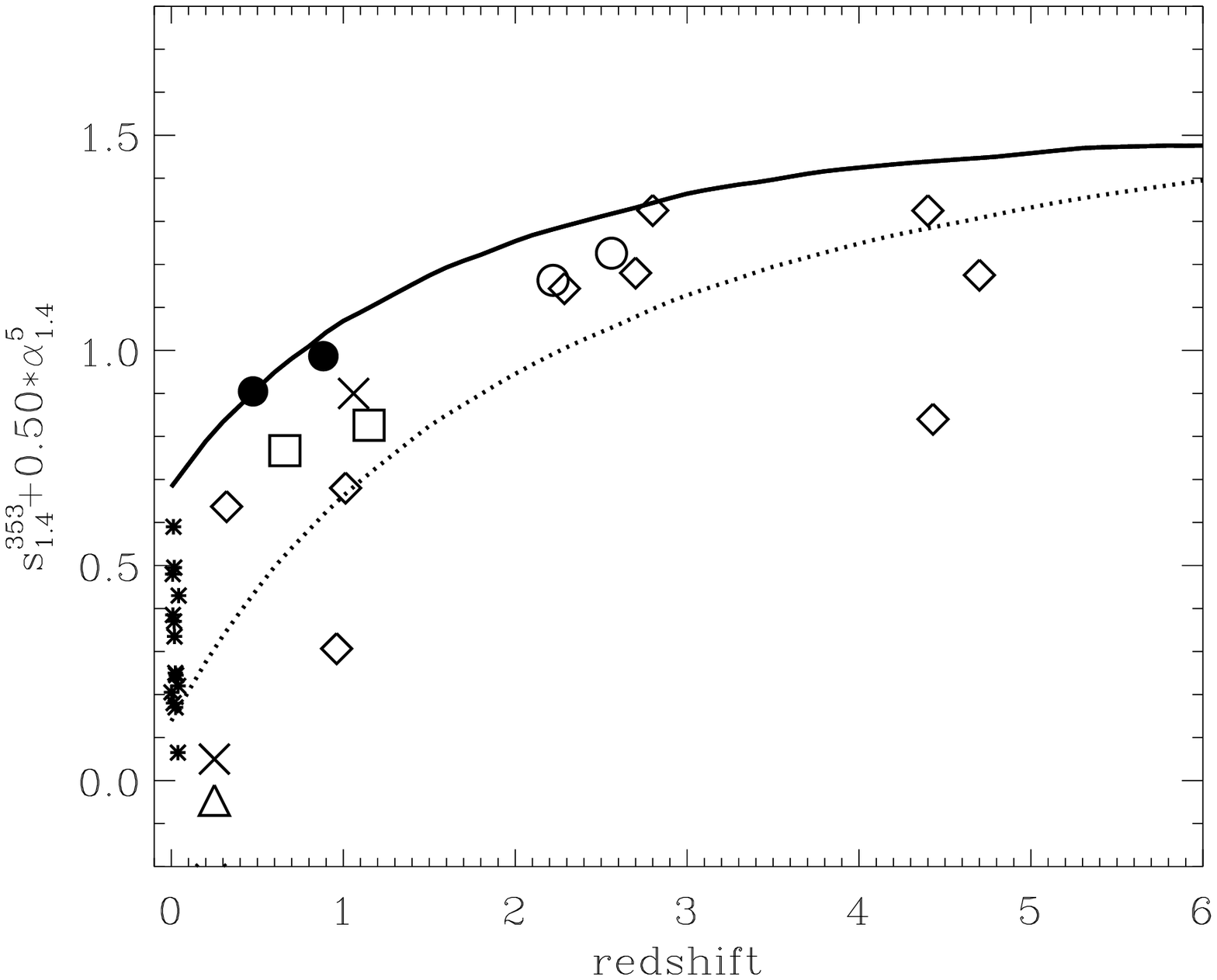}
\includegraphics[width=8truecm]{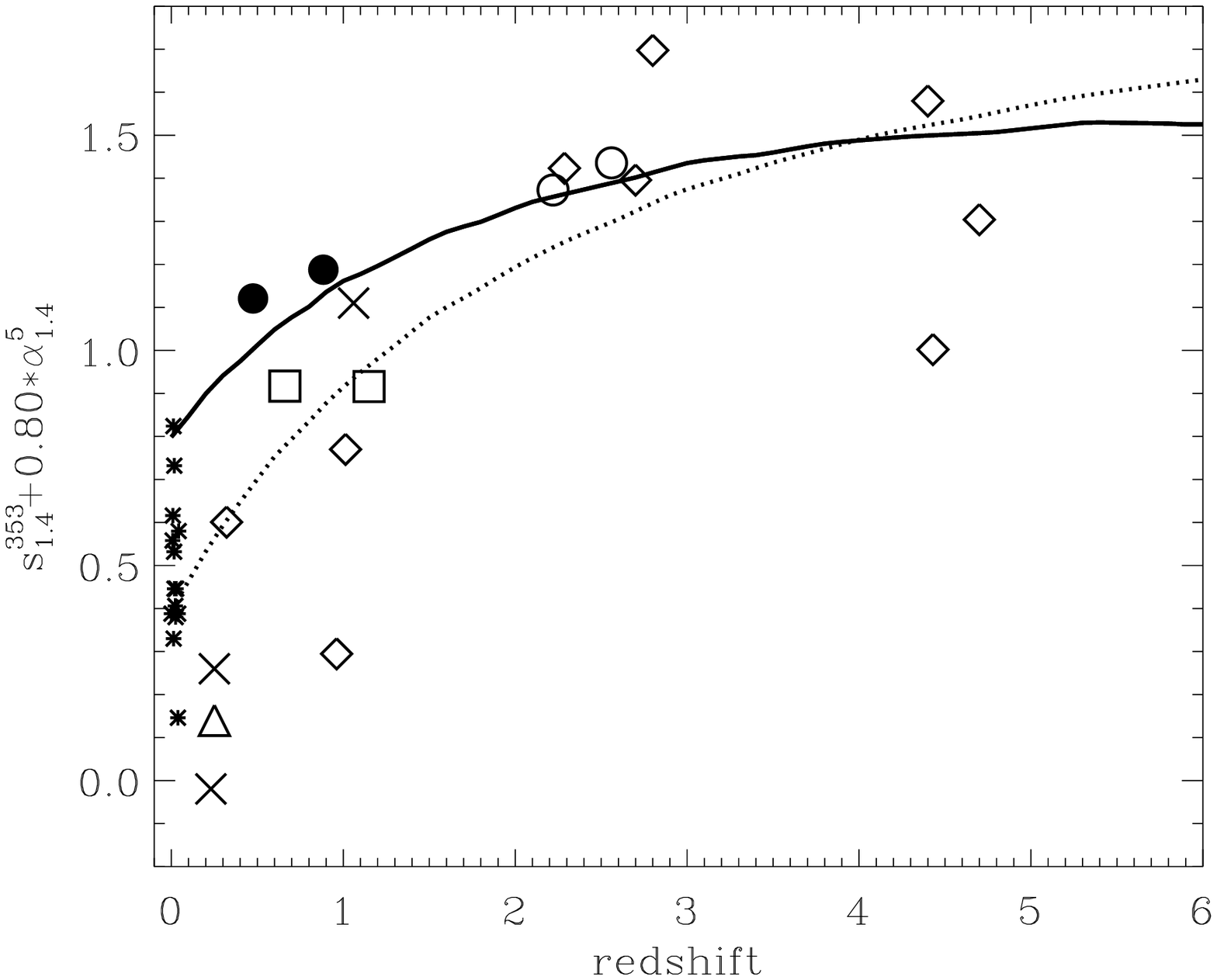}
\caption{Linear combinations of $s^{353}_{1.4}$ and $\alpha^{5}_{1.4}$ for models and data.
For data with only one radio flux available, the radio spectral index was set to 0.7.
See Fig.~\ref{alfa_z35314} for references to data.}
\label{alfa_zcomb}
\end{center}
\end{figure}

In Fig.~\ref{alfa_zcomb} we show that a linear combination of the
spectral index $s^{353}_{1.4}$ with a radio spectral index
($5-1.4$ GHz in the figures) may indeed reduce the redshift
uncertainty at each z originating from age dispersion. As an
example, $\Delta z$ at $z=2$ for models is 3.12, 1.67, 0.82
respectively for Fig. \ref{alfa_z35314} and the two panels of
Fig. \ref{alfa_zcomb}. On the other hand it is important to note
that, by considering the few available sources (excluding the
known AGN) with a spectroscopic z, the redshift range that would
correspond to their ordinate value depends on the adopted
spectral index or combination. A detailed investigation of the
redshift distribution of sub-mm galaxies is beyond the scope of
this paper.

\section{Discussion and conclusions}
\label{sconc}

In this paper we have revisited the nature of the FIR/Radio
correlation observed in star forming galaxies. To understand its
origin and range of validity we have utilized models of normal
star forming and starburst galaxies. The  infrared emission has
been estimated with our population synthesis code, GRASIL, which
is particularly suited for the prediction of the SED of star
forming galaxies, from the UV to the sub-mm. As for the radio
emission we have adopted a model which extends the capabilities
of GRASIL into the radio regime, essentially following the
prescriptions given by Condon \& Yin (1990). Before adopting it we
have reviewed all the possible sources of radio emission related
to the star formation activity, with particular emphasis on the
integrated properties of stellar populations.

In agreement with previous studies, we have found that the
fraction of the NT emission due to radio supernova remnants is
about 6\% of the total and that other discrete sources provide a
negligible contribution. Almost all the NT emission thus come
from diffuse electrons possibly injected into the interstellar
medium by CCSN events, and adopting a relation between the NT
emission and the  CCSN rate seems the safest way to proceed.
However, contrary to previous studies, we have also shown that the
latter relation, which is at the base of the tightness of the
FIR/Radio correlation, is a natural outcome {\it whenever
synchrotron losses happen on timescales that are shorter than the
fading time of the CCSN rate}.

This is certainly the case in normal spirals, where the global
SFR changes very little over the last billion
years. But the situation may be very different in luminous
starbursts, where the SN rate may change significantly in a
timescale typical of the lifetime of the most massive stars.
Indeed, previous studies claimed that the existence of a FIR/Radio
correlation under such conditions requires a fine tuning between
the magnetic and radiative energy density, which is difficult to
explain. We have shown that this {\it fine tuning} is not
necessary. In fact the observed prevalence of synchrotron energy
losses on inverse Compton against the photons of the high stellar
radiation field, indicates large magnetic fields and,
consequently, guarantees {\it very short synchrotron electron
lifetimes} (Condon 1992), certainly shorter than the typical
lifetime of the most massive stars. The proportionality between
the NT emission and the CCSN rate thus holds even in the extreme
conditions found in the luminous obscured compact starbursts.
This is why the FIR/Radio correlation appears so robust.

We have calibrated the NT emission/CCSN rate relation on the
observed properties of our Galaxy. With these assumptions {\it we
reproduce well the FIR/Radio correlation of normal star forming
galaxies}, namely q$_{1.4GHz}$=2.3. We are thus quite confident
that our model is able to reproduce both {\it the FIR and Radio
emission} of star forming galaxies, with a minimum set of well
calibrated parameters.

As for the starburst galaxies, with a rapidly changing SFR, we
have shown that the different fading times of the FIR and Radio
emission may be used to analyse in great detail the recent star
formation history in these galaxies. Indeed, in the cases of M82
and ARP220, presented in Sect.~\ref{sq}, the combination of FIR
and Radio observations allows to reach a time resolution of a few
tens of Myr, which is not possible based on the UV-FIR continuum
properties alone, and even on the optical/NIR emission lines
diagnostics, for obscured galaxies (Poggianti et al. 2001,
Bressan et al.\ 2001).
We have thus analysed in greater detail the evolution of obscured
starburst galaxies under different assumptions concerning the
burst characteristics, challenged by the possibility of using the
deviations from the FIR/Radio correlation as a diagnostic tool to
infer the recent star formation history. We have compared our
results with observations of a sample of compact ULIRGs, having
in mind that, in these objects, it is also not clear what
fraction of IR and radio emission is possibly contributed by
the AGN. Compact ULIRGs show a
noticeable deviation from the average FIR/Radio relation, with
their  q$_{1.49{\rm GHz}}$ being generally lager than 2.3, a few of
them with values as high as 3. Taken at face values, these
deviations suggest that radio emission is depressed by a large
factor, relatively to normal spirals. Understanding the interplay
between FIR and radio emission in these objects is thus
fundamental to make reliable predictions for high redshift dust
enshrouded galaxies.

Starburst models with peak SFR reaching several hundred
M$_\odot$/yr and thereafter exponentially declining, may account
for the IR and radio emission of the observed ULIRGs and are able
to reproduce the observed variation of the value of q$_{1.49{\rm GHz}}$.
This view is consistent with the current idea that ULIRGs and, to
a larger extent high redshift dust enshrouded galaxies, are
transient phenomena that nevertheless build up a significant
fraction of stars and metals (Granato et al. 2001).

The introduction of a new diagnostic diagram, the q$_{1.49{\rm GHz}}$
vs radio slope diagram, allows us to single out the effects of
starburst evolution and free-free absorption. Very young star
bursts display an excess of FIR emission relative to the radio
emission  because the latter is initially contributed mainly by
the free-free emission process. As the starburst ages, the NT
contribution increases and becomes the dominant source, while the
radio slope reaches the typical values observed in synchrotron
emission. Free-free absorption affects the 1.49 GHz data,
introducing a trend with the higher q being accompanied by the
shallower slope. The estimated optical depths for free-free
absorption at 1.49 GHz are between 0.5 and 1. At 8.4 GHz,
free-free absorption becomes negligible and the above trend
disappears. The value of q$_{8.4{\rm GHz}}$ is a measure of the age of
the starburst. However, even in the latter diagram the slope is
still affected by free-free absorption. Thus we suggest that a
similar diagram between 8.4 GHz and a higher frequency range
would be critical for the understanding of the evolutionary
status of  compact ULIRGs because, in that case, the slope,
unaffected by free-free absorption, would provide an independent
estimate of the age.

If ULIRGs are transient phenomena as suggested by other
independent studies, then determining their SFR from conventional
estimators may be a problem. They are far from being in a
stationary status, the term "average star formation" is
meaningless, and applying standard calibrations may result in a
significant error and/or apparent discrepancy between the
observable themselves. One should be able to reconstruct the
recent history of star formation and, for this purpose, we suggest
the use of the above diagram to determine the characteristic
parameters of the burst first, and then the age-averaged SFR
from either the FIR or radio luminosity.

Another relevant question addressed is how reliable is the use
the FIR/Radio correlation to evaluate the contribution of non
thermal radiation from the central active nucleus. Among the
plotted data, the symbol "M" indicates the position of the
Seyfert 1 galaxy  Mrk~231 (UGC 08058) (Thean et al. 2000). The
fact that Mrk~231 is clearly distinct from the other objects and
occupies a position below any starburst model, becomes
particularly evident in the q$_{8.4{\rm GHz}}$ vs slope diagram, where
the effects of free-free absorption on the q ratio are minimized.

We have also shown that during the {\it post starburst} phase,
the models reach values of q significantly lower than those of
quiescent spirals, with still significant FIR luminosities. This
is consistent with the detection in nearby Abell clusters of a
statistically significant excess of star forming galaxies with
enhanced radio emission relative to the FIR (Miller \& Owen
2001). We suggest that these low values of q are due to an
evolutionary effect rather than a direct enhancement of radio
emission by interaction with the intracluster medium.

Finally we have investigated on the redshift dependence of the
the FIR/Radio correlation and on its validity (through the sub-mm
radio index, $s^{353}_{1.4}$) to provide a photometric redshift
estimate of obscured distant galaxies. The large dispersion  of
$s^{353}_{1.4}$ observed among local galaxies is compatible with
the evolutionary effects discussed in Sect.~\ref{sq}. The unknown
evolutionary status of the starburst renders the $s^{353}_{1.4}$
index very unreliable at almost any redshift. We thus suggest to
complement the index $s^{353}_{1.4}$ with a radio slope
determination, because of its tight relation with the evolutionary
phase of the starburst and the its very shallow dependence on the
redshift. Other uncertainties like the sub-mm slope and the
presence of a significant contribution at radio wavelengths from
a central AGN, obviously worsen the above picture.

\begin{acknowledgements}
We thank I. Aretxaga , L. Danese, A. Franceschini, D. Huges, P. Panuzzo, B.
Poggianti and  O. Prouton  for useful discussions
and the anonymous referee for useful comments.
A.B. and G.L.G. acknowledge warm hospitality by INAOE. 
This research was
partially supported by the European Community under TMR grant
ERBFMRX-CT96-0086 and by 
the Italian Ministry
for University and Research (MURST) under grant Cofin 92001021149-002.
\end{acknowledgements}

\end{document}